\documentclass[
aps, 
prl,
preprintnumbers,
twocolumn,
superscriptaddress,
nofootinbib,
floatfix,
10pt
]{revtex4-2}

\usepackage{graphicx}
\usepackage{dcolumn}
\usepackage{bm}
\usepackage{braket}
\usepackage{amsmath}
\usepackage{amssymb}
\usepackage{mathtools}
\usepackage{comment}
\usepackage[bottom]{footmisc}
\usepackage{hyperref}

\def\orcid#1{\kern .08em\href{https://orcid.org/#1}{\includegraphics[keepaspectratio,width=0.7em]{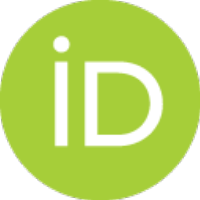}}}

\begin{document}

\title{\textbf{Diagrammatic Monte Carlo for Finite Systems at Zero Temperature} 
}%

\author{Stefano~Brolli\orcid{0009-0005-8290-8462}}
 \email{Stefano.Brolli@unimi.it}
\affiliation{Dipartimento di Fisica “Aldo Pontremoli,” Universit\`a degli Studi di Milano, via Celoria 16, I-20133 Milano, Italy}
\affiliation{INFN, Sezione di Milano, via Celoria 16, I-20133 Milano, Italy}

\author{Carlo~Barbieri\orcid{0000-0001-8658-6927}}
 \email{Carlo.Barbieri@unimi.it}
\affiliation{Dipartimento di Fisica “Aldo Pontremoli,” Universit\`a degli Studi di Milano, via Celoria 16, I-20133 Milano, Italy}
\affiliation{INFN, Sezione di Milano, via Celoria 16, I-20133 Milano, Italy}

\author{Enrico~Vigezzi\orcid{0000-0002-2683-6442}}
\email{Enrico.Vigezzi@mi.infn.it}
\affiliation{INFN, Sezione di Milano, via Celoria 16, I-20133 Milano, Italy}

\date{\today}

\begin{abstract}
\noindent High-order virtual excitations play an important role in microscopic models of nuclear reactions at intermediate energies. However, the factorial growth of their complexity has prevented their consistent inclusion in \emph{ab initio} many-body calculations.
For infinite systems at finite temperature, such drawbacks can be overcome using diagrammatic Monte Carlo techniques to resum entire series of Feynman diagrams. We present a diagrammatic Monte Carlo algorithm that can be applied to self-bound systems with discrete energy levels at zero temperature, and demonstrate its potential for the Richardson model of nuclear pairing.
We show that sampling the topological space of diagrams allows the inclusion of high-order excitations that are neglected in state-of-the-art approximations used in nuclear physics and quantum chemistry.
We propose that sampling the diagrammatic space can overcome the long-standing gap between our microscopic understanding of structure and reactions in nuclear physics.
\end{abstract}

\maketitle

\begin{figure}[b]
\centering
\includegraphics[width=0.48\textwidth]
{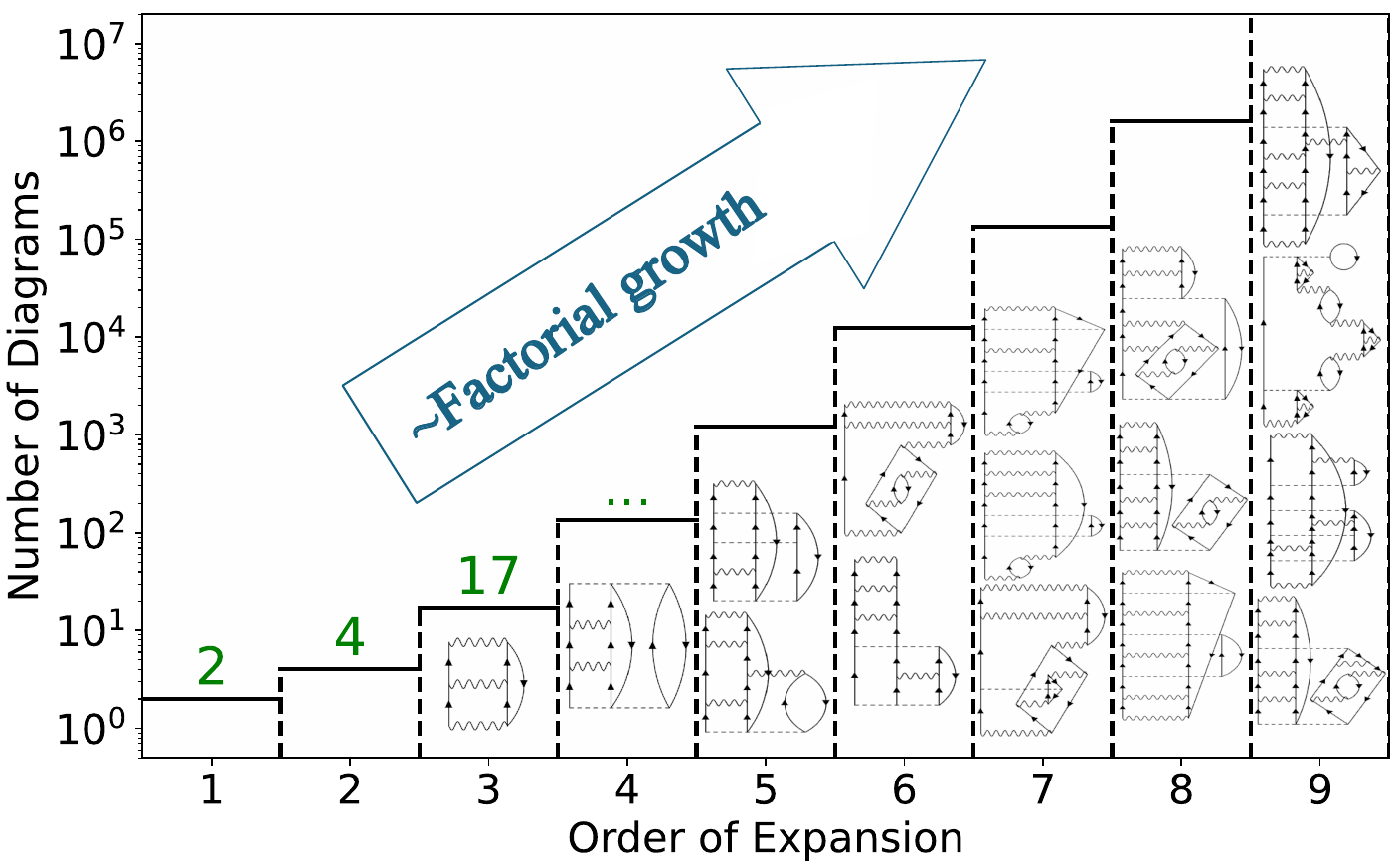}
\caption{Approximate number of Feynman diagrams contributing to the self-energy for two- and three-body forces. The values shown in green give the exact number of Feynman diagrams for the first three orders~\cite{Raimondi2018} while values at higher order are estimated based on their factorial scaling. For each order, very few representative diagrams are shown.}
\label{figFeynDiags}
\end{figure}

\emph{Ab initio} nuclear theory has significantly advanced over the past two decades by combining effective field theories (EFTs) of the nuclear force with the latest computational approaches to the quantum many-body problem.
Several high-precision methods can predict the ground state properties of medium-light nuclei~\cite{Coraggio2021,Hergert2020}, and recently reached heavy elements~\cite{Arthuis2020, Hu2022}. The EFT interactions used in modern
literature are rooted in the symmetries of QCD and avoid relying on the phenomenology of stable nuclei, so that controlled and accurate predictions remain feasible even toward the limits of nuclear existence. Despite these successes, the microscopic prediction of reaction cross sections, as measured by advanced radioactive ion beam facilities~\cite{RIBF,FRIB,FAIR,ISOLDE,GANIL,ARIEL,SPES}, is still an unsettled challenge for \emph{ab initio} theory. Exotic neutron-rich nuclei are essential for studying key physics questions, including shell evolution~\cite{Otsuka2021, Tsunoda2020}, heavy element formation~\cite{Ravlic2023}, the role of the r process during neutron star mergers~\cite{Cowan2021,Johnson2019}, and many others. 
Among the available \emph{ab initio} methods~\cite{Baroni2013,Rotureau2018,Idini2019,Burrows2024,Vorabbi2024}, the self-consistent Green’s function approach is suited for a unified description of structure and reactions. The spectral distribution, which gathers all one-nucleon addition and removal information, the ground-state observables, and the exact microscopic optical potential are all consistently included in the formalism~\cite{Capuzzi1996, Cederbaum2001, Escher2002, Dickhoff:manybody}. The Green's function $G$ is the exact solution of the Dyson equation $G_{\alpha \beta} (\omega) = G^{(0)}_{\alpha \beta} (\omega) + \sum_{\gamma \delta} G^{(0)}_{\alpha \gamma} (\omega) \Sigma^{\star}_{\gamma \delta} (\omega) G_{\delta \beta} (\omega)$, where $G^{(0)}$ describes a freely propagating particle (or hole) and the term $G^{(0)} \Sigma^{\star} G$ introduces recursive dynamic interactions with the other particles of the system.
In this picture, the self-energy $\Sigma^{\star}$ is a complex, nonlocal, energy-dependent microscopic optical potential~\cite{Mahaux1995, Capuzzi1996, Cederbaum2001}. Self-consistent Green's function theory relies on systematically improvable many-body truncations of the self-energy~\cite{Barbieri::LectNotesPhys936}; however, current approximation schemes do not include high-order configurations that dominate scattering in the mid-energy range of 10-100 MeV~\cite{Idini2019, Waldecker2011}.
Because of the factorial growth of Feynman diagrams (see Fig. \ref{figFeynDiags}), improving the accuracy of \emph{ab initio} optical potentials requires going beyond the order-by-order addition of terms and instead aiming for a stochastic resummation by sampling a limited number of representative higher-order diagrams.

Sampling different diagram topologies is the fundamental idea of the diagrammatic Monte Carlo (DiagMC) approach~\cite{Prokofev1998}. This method first gained much interest in the field of condensed matter physics, where it was employed to study infinite systems at finite temperature such as the polaron~\cite{Prokofev2008} and the unitary Fermi gas~\cite{VanHoucke2019, VanHoucke2012}. More recent advances in the DiagMC framework have been applied to study single-particle excitations in the electron gas~\cite{Haule2022, Chen2019} and the real-time dynamics of dissipative quantum impurity models~\cite{Vanhoecke2024}.
To our knowledge, the DiagMC approach has still been poorly investigated for bound systems at zero temperature such as atomic nuclei, where (large) discrete bases are usually needed. Furthermore, applications to scattering theory require satisfying the causality principle, which is encoded in a dispersive spectral relation. This property is naturally broken by sharp truncations in the Feynman perturbation series, and it has never been enforced in previous DiagMC calculations.
In this Letter, we introduce a DiagMC algorithm for the Richardson model and impose the correct dispersive relation by representing the stochastic self-energy with a parametrization suitable for finite systems. 

\paragraph {\itshape The Richardson model} --- The Richardson model was introduced as a simplified description of nuclear superfluidity. It explains pairing correlations in Sn and Pb isotopes with considerable accuracy~\cite{Baerdemacker2014Sn, Richardson:EvenLead}, as well as in quantum chemistry systems~\cite{Johnson2020}. 
Furthermore, it provides stringent tests for quantum many-body theories that aim to describe strongly correlated Fermi liquids~\cite{Hjorth-Jensen::LectNotesPhys936}, and has challenged quantum neural networks~\cite{Rigo2023} and eigenvector continuation~\cite{Franzke2024}. The availability of exact solutions makes it suitable for precise benchmarks.

The Richardson Hamiltonian includes a contact interaction between nucleons that occupy the same energy level~\cite{Richardson:Eigenstates}. 
We consider $D$ doubly degenerate equidistant energy levels and we focus on the half-filled system with $A = 2P$ fermions,  
\begin{equation} \label{eqRichH}
H^{(D)} = \xi \sum_{p = 1}^{D} \sum_{s = \uparrow, \downarrow} \left( p - 1 \right) c_{p s}^{\dagger} c_{p s} - \frac{g}{2} \sum_{p, q = 1}^{D} c_{p \uparrow}^{\dagger} c_{p \downarrow}^{\dagger} c_{q \downarrow} c_{q \uparrow},
\end{equation}
where we set $\xi = 1$ without loss of generality.
The ground state of this Hamiltonian can be written analytically as the product of the $P$ lowest correlated-pair creation operators, $b^\dagger_i$, such that $\ket{\Psi^A_0} = \prod_{i = 1}^{P} b_{i}^{\dagger} \ket{0}$~\cite{SupplMat}.

\noindent The propagator theory of the Richardson model was explored in Refs.~\cite{Barbieri::LectNotesPhys936} and~\cite{Brolli::Masters}. The one-body propagator is diagonal and invariant under spin exchange, and hence only its $D$ diagonal and spin-up components need to be considered. Furthermore, we found that ladder diagrams greatly dominate the self-energy expansion of the propagator in the perturbative regime. We will focus on this subset of diagrams in our sampling and show that this is sufficient to compute the full correlation energy within the stochastic error and the error induced by a finite regulator.

\begin{figure}[t]
    \centering
    \begin{minipage}[t]{0.48\textwidth}
        \raggedright
        \textbf{a)} \\ 
        \includegraphics[width=\textwidth]{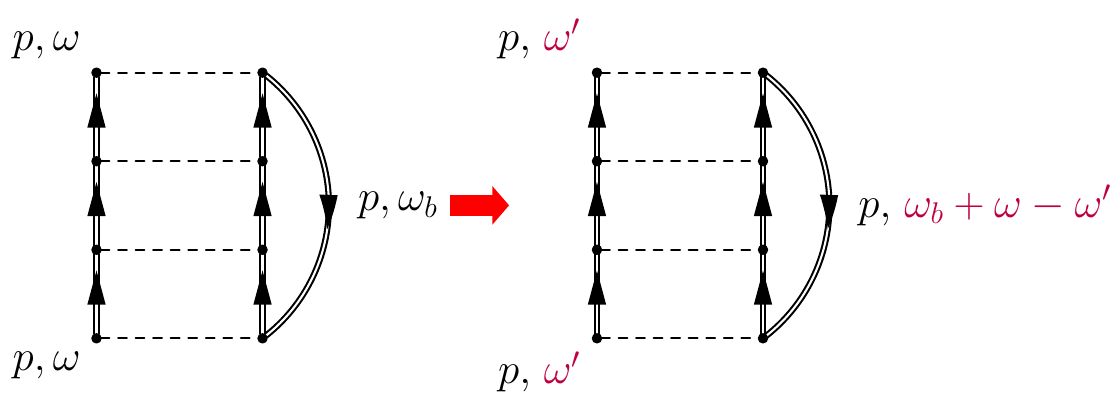}
    \end{minipage}%
    \hfill
    \begin{minipage}[t]{0.48\textwidth}
        \raggedright
        \textbf{b)} \\ 
        \includegraphics[width=\textwidth]{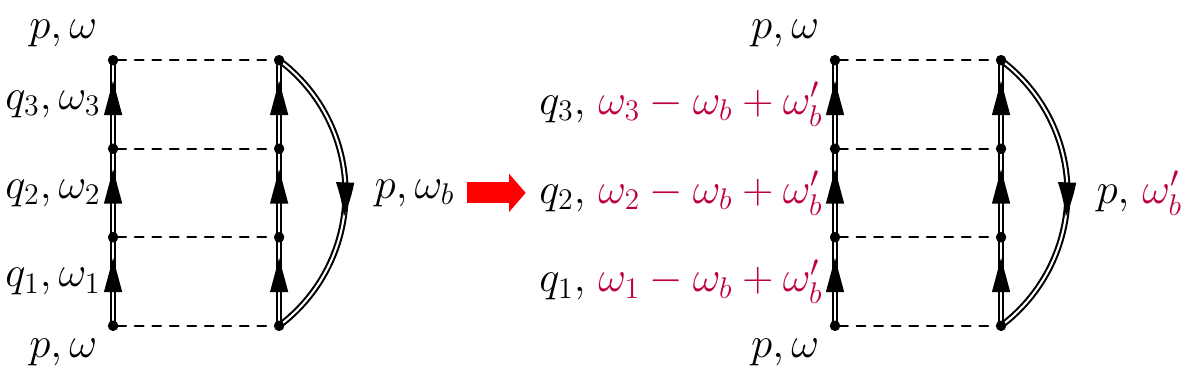}
    \end{minipage}
    
    \vskip\baselineskip
    
    \begin{minipage}[t]{0.48\textwidth}
        \raggedright
        \textbf{c)} \\ 
        \includegraphics[width=\textwidth]{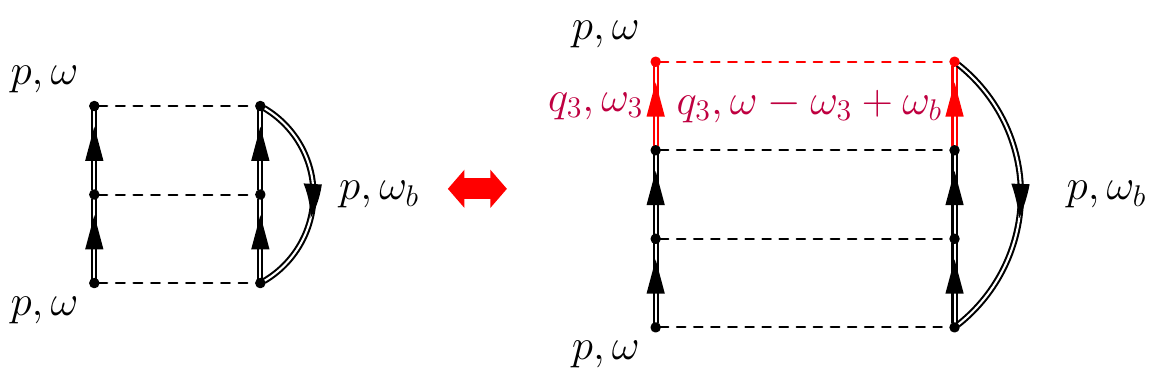}
    \end{minipage}%
    \hfill
    \begin{minipage}[t]{0.48\textwidth}
        \raggedright
        \textbf{d)} \\ 
        \includegraphics[width=\textwidth]{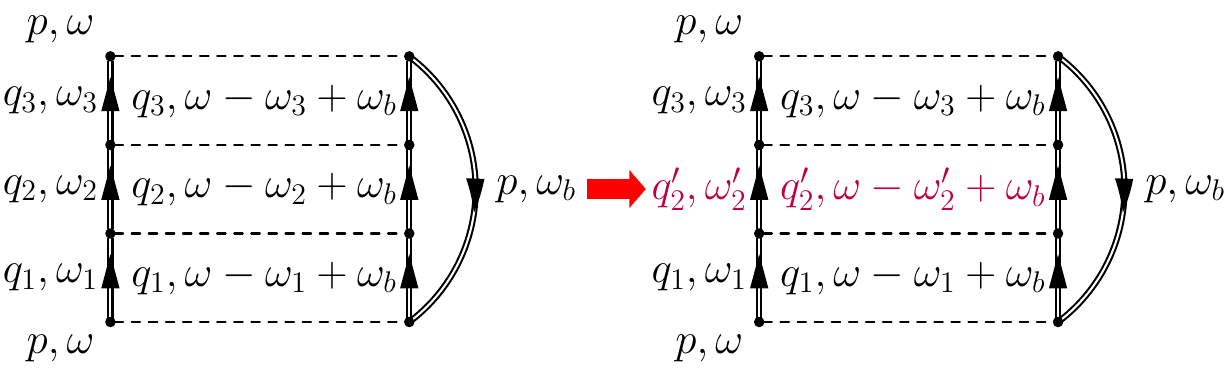}
    \end{minipage}
    \caption{Updates used to sample the entire ladder diagrammatic expansion \cite{SupplMat}.
    Rules (a), (b), (d) yield the standard Monte Carlo evaluation of single diagrams. Rule (c) allows to span the topological space of all self-energy ladder diagrams.}
    \label{figMCRules}
\end{figure}

\paragraph{\itshape Diagrammatic Monte Carlo} ---
Standard Monte Carlo integration has been used to compute many-body perturbation theory (PT) for nucleonic matter reaching up to the full fourth order~\cite{Drischler2019} but relies on the cumbersome determination and evaluation of every single PT contribution~\cite{ArthuisADG1_2019}. 
The DiagMC framework used in this work overcomes this limitation by promoting diagram topologies on the same footing as internal quantum numbers and frequencies ~\cite{VanHoucke2019, Prokofev1998}. 
We express the exact self-energy according to the fundamental identity
\begin{equation} \label{eqDiagMCMainText}
\Sigma_{p}^{\star} (\omega) = Z_p \int d \mathcal{C} \frac{\lvert \mathcal{D}_p\left( \omega; \mathcal{C} \right) \rvert}{Z_p} \exp \left[ i \arg\mathcal{D}_p \left( \omega; \mathcal{C} \right) \right],
\end{equation}
where $\mathcal{D}_p$ is a diagram characterized by its topology and internal single-particle quantum numbers and frequencies, which we collectively label $\mathcal{C}$, and $Z_p$ is a normalization factor. The integral over $d \mathcal{C}$ includes the sum over all topologies. If the diagrams are sampled according to a probability distribution $\propto \lvert \mathcal{D}_p\left( \omega; \mathcal{C} \right) \rvert$, the self-energy can be calculated as a stochastic sum of the phases of the diagrams.
The sampling of ladder diagrams is implemented through a Markov chain in the space of diagrams. At each step of the chain, one of the updates in Fig. \ref{figMCRules} is called at random, and a walker in $\mathcal{C}$ space is generated using the Metropolis-Hastings algorithm~\cite{Metropolis1953, Hasting1970}.
We refer to the Supplemental Material for a comprehensive overview of the present DiagMC implementation~\cite{SupplMat}.

The DiagMC algorithm is built upon the Feynman perturbation expansion which, if truncated sharply at each order, can break important nonperturbative features of the exact self-energy. Most notably, the causality principle that is important for scattering events may require full all-order summations. 
Causality is expressed through the boundary condition of the K\"{a}llén-Lehmann representation for the exact self-energy:

\begin{equation} \label{eqLeSE}
\Sigma_{p \uparrow}^{\star} \left( \omega \right) = \Sigma_{p}^{(\infty)} + \sum_{n = 1} \frac{A_{p}^{n}}{\omega - B_{p}^n + i \eta} + \sum_{k = 1} \frac{A_{p}^{k}}{\omega - B_{p}^k - i \eta},
\end{equation}

\noindent where $A_p^{n/k} > 0$ and $B_p^{n/k}$ are, respectively, the strength and energy position of the $n/k$ th pole, and must be determined by the DiagMC simulation. Note that the dispersion relation implies that only the static mean-field term, $\Sigma_{p}^{(\infty)}$, and the imaginary part of \eqref{eqLeSE} are required to reconstruct the entire self-energy~\cite{Dickhoff:manybody}. The algebraic diagrammatic construction (ADC) method that is state of the art for nuclear physics and quantum chemistry was developed with the specific aim of retaining this structure exactly~\cite{Schirmer2018LNC, Barbieri::LectNotesPhys936, Schirmer1983::ADC}.

The $\Sigma_{p}^{(\infty)}$ term is a tadpole diagram that plays a special role in the DiagMC algorithm. Its expression includes a convergence factor $e^{i\omega\eta}$ that makes the Monte Carlo integration inefficient; however, it can be easily calculated with standard numerical methods.
Moreover, the DiagMC algorithm calculates the self-energy only up to the overall (real) factor $Z_p$ of Eq.~\eqref{eqDiagMCMainText}. To fix this ambiguity and at the same time address the convergence problem of the tadpole diagrams, we approached the problem in a novel way. First, the $\Sigma_{p}^{(\infty)}$ diagram, and any diagram that includes repetitions of it, are simple enough that they can be evaluated exactly with an analytical frequency integration followed by standard integration over the remaining single-particle quantum numbers. Hence, we compute it separately to avoid any Monte Carlo instability.
Second, the Markov chain walker of the DiagMC algorithm can still be implemented to retain all steps over the tadpole diagrams: simply such contributions are ignored in the sum of Eq.~\eqref{eqDiagMCMainText} to avoid any double counting. This fact gives us the additional freedom needed to determine the overall normalization. We replace the tadpole diagrams with unphysical ones that have an exactly known analytical solution and compare it with the partial DiagMC resummation over the same terms. We can deduce the overall normalization factor by counting how many times the unphysical diagram is visited compared to the total number of samples~\cite{SupplMat}.
\begin{figure}[!t]
\includegraphics[width=0.48\textwidth]{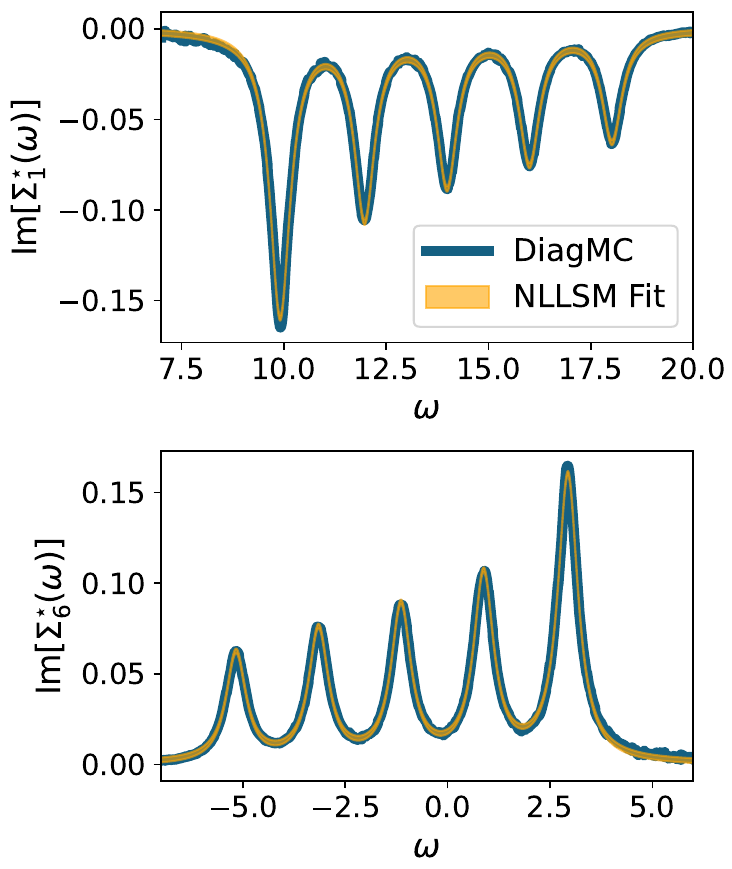}
\caption{Imaginary part of the diagonal components $p = 1, 6$ of the irreducible self-energy. The self-energy was calculated with the DiagMC algorithm and fitted via NLLSM with a sum of Lorentzian functions. The simulation was performed with a HF reference propagator in a model space of dimension $D = 10$ and $A = 10$ particles. We used $g = 0.3$, $\eta = 0.1$. DiagMC sampled the ladder series up to the eighth order. The Lorentzian representation is very accurate due to the small stochastic noise, which is barely visible only in the far tails.}
\label{figSE}
\end{figure}

Within the radius of convergence of the perturbation expansion ($|g|<g_c$), increasing the truncation order should bring the simulation closer to the correct Källén-Lehmann representation. This radius depends on the regulator $\eta$, and increases for lower resolution (i.e., larger values of $\eta$). Choosing $\eta = 0.1$ ensures effectively approaching the limit $\eta \to 0$, and we determined $g_c\approx 0.3$ in this case~\cite{SupplMat}. An important novelty in our approach is to perform a nonlinear least-square minimization (NLLSM) that fits the imaginary part of the self-energy calculated with the DiagMC algorithm as a sum of Lorentzians. Dispersion relations are used to recover a natural parameterization of the full self-energy that fulfills the causality principle. This approach allows one to control the Monte Carlo noise of the simulation by translating it into a (small) error on the fitted parameters~\cite{SupplMat}.

\paragraph {\itshape Results} --- We perform calculations in the so-called $sc0$ framework~\cite{Soma2014::Gorkov, Barbieri2022::GorkovADC(3)}. This is a standard prescription in nuclear theory that consists of imposing a fully self-consistent calculation for the static component of the self-energy, $\Sigma_{p}^{(\infty)}$, while the dynamic part is expanded with respect to a mean-field reference propagator. 
Figure \ref{figSE} reports the components $p = 1$ and $p = 6$ of the imaginary part of the irreducible self-energy for $g = 0.3$, $\eta = 0.1$, and $D=10$. The results were obtained with a Hartree-Fock (HF) reference propagator by sampling ladder diagrams up to order eight. The simulated self-energy shows only small deviations from the exact analytic form, Eq.~\eqref{eqLeSE}, so that the NLLSM fit is extremely precise.

Our implementation of the DiagMC approach carries three sources of uncertainty.
The first is the statistical error that is intrinsic to any Monte Carlo approach and can be controlled to a good extent by additional sampling. 
Second, we do perform a many-body truncation by restricting our diagram space up to eighth order in PT and to ladder diagrams only. The error involved is negligible because excluded diagram topologies are known to be irrelevant for the Richardson Hamiltonian~\cite{Barbieri::LectNotesPhys936} and the first eight PT orders are substantially converged~\cite{SupplMat}. The comparison between the NLLSM fit of Eq.~\eqref{eqLeSE} and the DiagMC sampling, shown in Fig.~\ref{figSE}, demonstrates the quality of our simulations.
Finally, we have a dependence on the regulator parameter~$\eta$~\cite{SupplMat}. In principle, the exact result is recovered in the limit $\eta\to 0$, where it would cause numerical instabilities. In fact, however, a finite value of~$\eta$ simply sets the energy resolution of our simulations, and how ``small'' it should be depends on the specific application.
\begin{figure}[t]
\centering
\includegraphics[scale = .55]{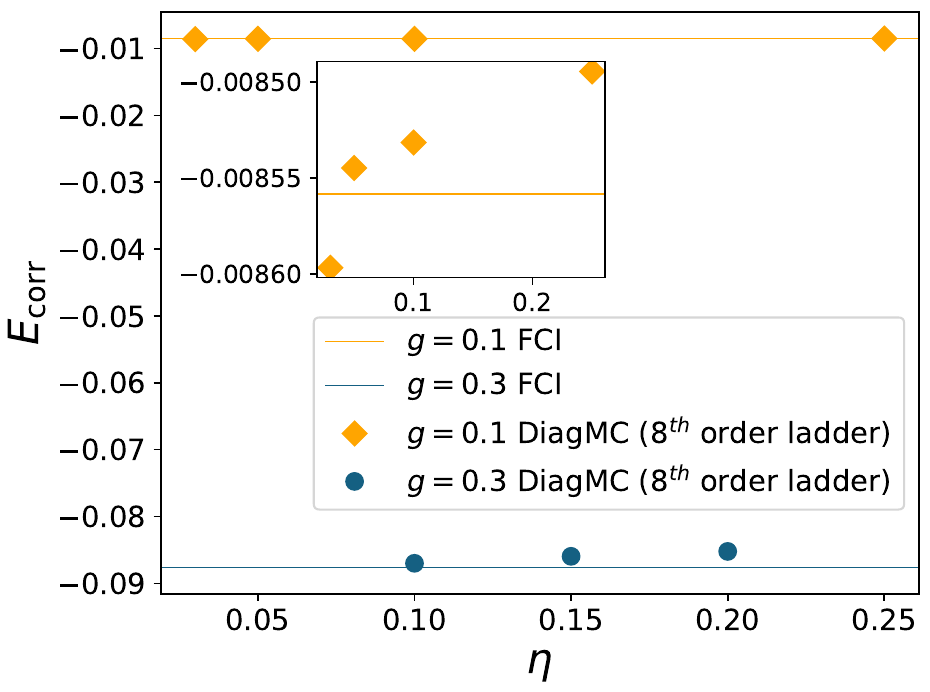}
\caption{Correlation energy for different values of the regulator $\eta$. The figure shows results for the coupling constants $g=0.1$ and $g = 0.3$. The solid lines are the full configuration interaction (FCI) results, and they are obtained from the exact diagonalization of the Richardson Hamiltonian.}
\label{figEcorrReg}
\end{figure}
To study the dependence of our results on $\eta$ we performed simulations using different regulator values in the model space $D=10$ for $g = 0.1$ and $g=0.3$.
Figure \ref{figEcorrReg} shows a moderate dependence on the regulator $\eta$. The deviation between our calculations and the true correlation energies is at most $1\%$ for both $g=0.1$ and $g=0.3$ when $\eta \leq 0.1$. Hence, we take $1 \%$ as the error for all our calculations of correlation energies and always use $\eta = 0.1$.

\begin{figure}[t]
\centering
\includegraphics[width=0.48\textwidth]{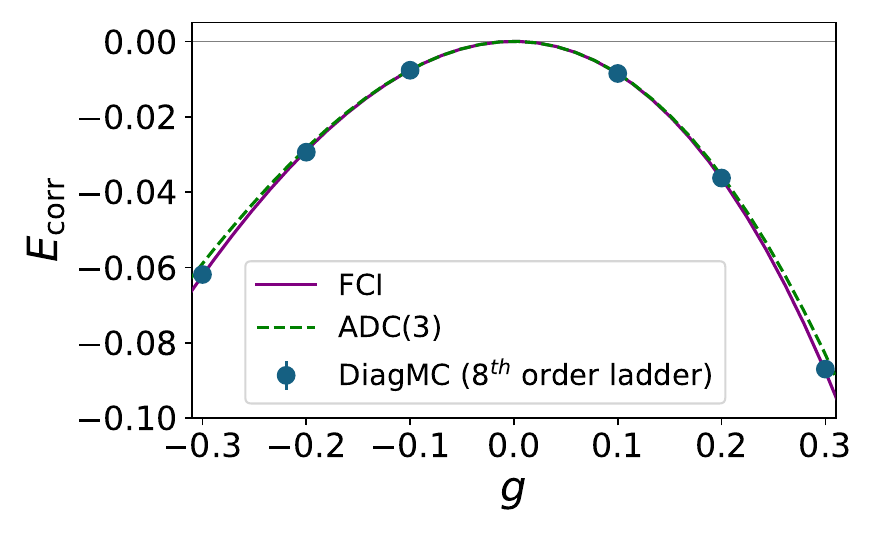}
\includegraphics[width=0.48\textwidth]{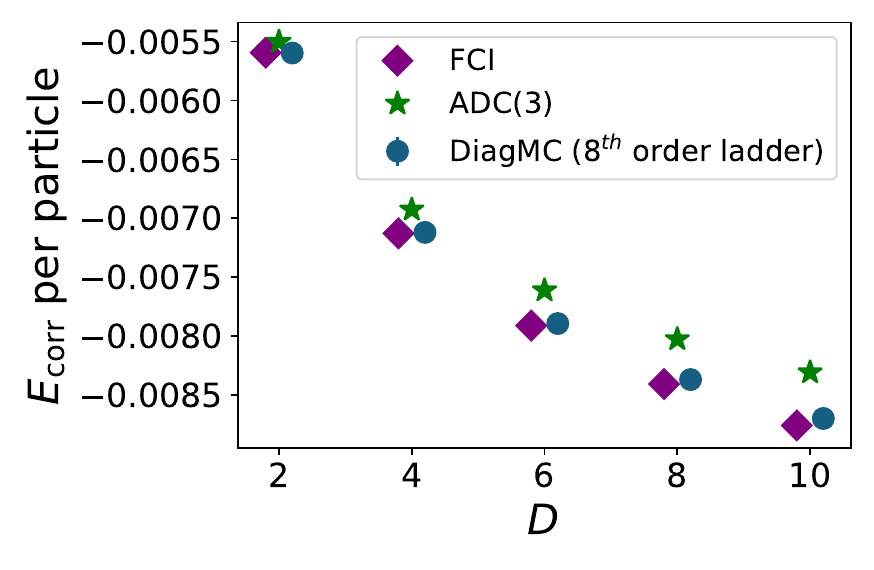}
\caption{Correlation energy of the Richardson model at half-filling. The regulator is set at $\eta = 0.1$ and the error of $1\%$ on the correlation energy is too small to be visible in the plot. We restricted our analysis to the values inside the convergence radius of the perturbation expansion, which we determined to be $g_c\approx 0.3$. Top panel: dependence of the correlation energy on the coupling constant $g$ in the model space $D=10$. Bottom panel: correlation energy per particle for different model space dimensions and $g = 0.3$.}
\label{figCorrE}
\end{figure}

The correlation energies derived from the Migdal-Galitskii-Koltun sum rule \cite{Migdal1958, Koltun1974} are benchmarked against the exact full configuration interaction (FCI) solutions and the ADC(3)  many-body truncation values in Fig.~\ref{figCorrE}. The top panel shows the correlation energy dependence on the coupling constant $g$ within the model space of size $D = 10$. The bottom panel shows the correlation energy per particle with changing model space size $D$, at half-filling, and $g = 0.3$. 
We find agreement with the exact energies within the error bars of the Monte Carlo uncertainty, confirming that the present ladder resummation dominates the self-energy expansion.
Remarkably, the DiagMC algorithm with the HF reference propagator outperforms ADC(3). This discrepancy is to be ascribed to the presence of time inversion terms in the ladder series, which is resummed to all orders in ADC(3) but only limited to a Tamm-Dancoff resummation~\cite{Barbieri::LectNotesPhys936}. The DiagMC algorithm correctly resums the Feynman ladders, accounting for all possible time orderings.

{\itshape Conclusion} --- For the weak-coupling regime, we conclude that the DiagMC approach on bound systems (with discrete spectra) can match and even supersede state-of-the-art methods for computing many-body propagators, including the ADC(3) truncation scheme employed in \emph{ab initio} nuclear physics and quantum chemistry.
The Richardson model could be solved without the need to exploit convergence acceleration methods, such as Borel resummation, that have already been widely used in DiagMC applications to condensed matter~\cite{VanHoucke2010, Rossi2018}. This leaves room for approaching more realistic problems.
The self-energy of realistic nuclear Hamiltonians is not dominated by the ladder series, so that new sets of updates that can sample any relevant diagram will be needed. A possible strategy is to adopt already existing schemes~\cite{VanHoucke2019} and integrate them with additional update rules aimed at improving the sampling in self-bound systems.
This approach appears to be the most straightforward way to tackle realistic nuclear systems.
It should be stressed that this work focused on the Green's function; however, the DiagMC approach can be extended to other methods based on diagrammatic expressions, such as coupled cluster amplitudes or many-body PT contributions. Therefore, the DiagMC technique holds significant potential for advancing various fields of physics involving bound many-fermion systems at zero temperature.  
The success of the present calculations signifies an initial step toward a unified description of nuclear structure and reactions by performing a stochastic sampling over the space of Feynman diagrams of a simplified pairing Hamiltonian.

\paragraph {\itshape Acknowledgments} --- This work used the DiRAC Data Intensive service (DIaL3) at the University of Leicester, managed by the University of Leicester Research Computing Service on behalf of the STFC DiRAC HPC Facility (www.dirac.ac.uk). The DiRAC service at Leicester was funded by BEIS, UKRI and STFC capital funding and STFC operations grants. DiRAC is part of the UKRI Digital Research Infrastructure.

\paragraph {\itshape Data availability} --- The data that support the findings of
this article are not publicly available. The data are available
from the authors upon reasonable request.

\bibliographystyle{apsrev4-2}
\bibliography{bibliography}

\begin{thebibliography}{57}%
\makeatletter
\providecommand \@ifxundefined [1]{%
 \@ifx{#1\undefined}
}%
\providecommand \@ifnum [1]{%
 \ifnum #1\expandafter \@firstoftwo
 \else \expandafter \@secondoftwo
 \fi
}%
\providecommand \@ifx [1]{%
 \ifx #1\expandafter \@firstoftwo
 \else \expandafter \@secondoftwo
 \fi
}%
\providecommand \natexlab [1]{#1}%
\providecommand \enquote  [1]{``#1''}%
\providecommand \bibnamefont  [1]{#1}%
\providecommand \bibfnamefont [1]{#1}%
\providecommand \citenamefont [1]{#1}%
\providecommand \href@noop [0]{\@secondoftwo}%
\providecommand \href [0]{\begingroup \@sanitize@url \@href}%
\providecommand \@href[1]{\@@startlink{#1}\@@href}%
\providecommand \@@href[1]{\endgroup#1\@@endlink}%
\providecommand \@sanitize@url [0]{\catcode `\\12\catcode `\$12\catcode
  `\&12\catcode `\#12\catcode `\^12\catcode `\_12\catcode `\%12\relax}%
\providecommand \@@startlink[1]{}%
\providecommand \@@endlink[0]{}%
\providecommand \url  [0]{\begingroup\@sanitize@url \@url }%
\providecommand \@url [1]{\endgroup\@href {#1}{\urlprefix }}%
\providecommand \urlprefix  [0]{URL }%
\providecommand \Eprint [0]{\href }%
\providecommand \doibase [0]{https://doi.org/}%
\providecommand \selectlanguage [0]{\@gobble}%
\providecommand \bibinfo  [0]{\@secondoftwo}%
\providecommand \bibfield  [0]{\@secondoftwo}%
\providecommand \translation [1]{[#1]}%
\providecommand \BibitemOpen [0]{}%
\providecommand \bibitemStop [0]{}%
\providecommand \bibitemNoStop [0]{.\EOS\space}%
\providecommand \EOS [0]{\spacefactor3000\relax}%
\providecommand \BibitemShut  [1]{\csname bibitem#1\endcsname}%
\let\auto@bib@innerbib\@empty
\bibitem [{\citenamefont {Coraggio}\ \emph {et~al.}(2021)\citenamefont
  {Coraggio}, \citenamefont {Pastore},\ and\ \citenamefont
  {Barbieri}}]{Coraggio2021}%
  \BibitemOpen
  \bibfield  {author} {\bibinfo {author} {\bibfnamefont {L.}~\bibnamefont
  {Coraggio}}, \bibinfo {author} {\bibfnamefont {S.}~\bibnamefont {Pastore}},\
  and\ \bibinfo {author} {\bibfnamefont {C.}~\bibnamefont {Barbieri}},\ }\href
  {https://doi.org/10.3389/fphy.2020.626976} {\bibfield  {journal} {\bibinfo
  {journal} {Front. Phys.}\ }\textbf {\bibinfo {volume} {8}},\ \bibinfo {pages}
  {626976} (\bibinfo {year} {2021})}\BibitemShut {NoStop}%
\bibitem [{\citenamefont {Hergert}(2020)}]{Hergert2020}%
  \BibitemOpen
  \bibfield  {author} {\bibinfo {author} {\bibfnamefont {H.}~\bibnamefont
  {Hergert}},\ }\href {https://doi.org/10.3389/fphy.2020.00379} {\bibfield
  {journal} {\bibinfo  {journal} {Front. Phys.}\ }\textbf {\bibinfo {volume}
  {8}},\ \bibinfo {pages} {379} (\bibinfo {year} {2020})}\BibitemShut {NoStop}%
\bibitem [{\citenamefont {Arthuis}\ \emph {et~al.}(2020)\citenamefont
  {Arthuis}, \citenamefont {Barbieri}, \citenamefont {Vorabbi},\ and\
  \citenamefont {Finelli}}]{Arthuis2020}%
  \BibitemOpen
  \bibfield  {author} {\bibinfo {author} {\bibfnamefont {P.}~\bibnamefont
  {Arthuis}}, \bibinfo {author} {\bibfnamefont {C.}~\bibnamefont {Barbieri}},
  \bibinfo {author} {\bibfnamefont {M.}~\bibnamefont {Vorabbi}},\ and\ \bibinfo
  {author} {\bibfnamefont {P.}~\bibnamefont {Finelli}},\ }\href
  {https://doi.org/10.1103/PhysRevLett.125.182501} {\bibfield  {journal}
  {\bibinfo  {journal} {Phys. Rev. Lett.}\ }\textbf {\bibinfo {volume} {125}},\
  \bibinfo {pages} {182501} (\bibinfo {year} {2020})}\BibitemShut {NoStop}%
\bibitem [{\citenamefont {Hu}\ \emph {et~al.}(2022)\citenamefont {Hu},
  \citenamefont {Jiang}, \citenamefont {Miyagi}, \citenamefont {Sun},
  \citenamefont {Ekstr{\"o}m}, \citenamefont {Forss{\'e}n}, \citenamefont
  {Hagen}, \citenamefont {Holt}, \citenamefont {Papenbrock}, \citenamefont
  {Stroberg},\ and\ \citenamefont {Vernon}}]{Hu2022}%
  \BibitemOpen
  \bibfield  {author} {\bibinfo {author} {\bibfnamefont {B.}~\bibnamefont
  {Hu}}, \bibinfo {author} {\bibfnamefont {W.}~\bibnamefont {Jiang}}, \bibinfo
  {author} {\bibfnamefont {T.}~\bibnamefont {Miyagi}}, \bibinfo {author}
  {\bibfnamefont {Z.}~\bibnamefont {Sun}}, \bibinfo {author} {\bibfnamefont
  {A.}~\bibnamefont {Ekstr{\"o}m}}, \bibinfo {author} {\bibfnamefont
  {C.}~\bibnamefont {Forss{\'e}n}}, \bibinfo {author} {\bibfnamefont
  {G.}~\bibnamefont {Hagen}}, \bibinfo {author} {\bibfnamefont {J.~D.}\
  \bibnamefont {Holt}}, \bibinfo {author} {\bibfnamefont {T.}~\bibnamefont
  {Papenbrock}}, \bibinfo {author} {\bibfnamefont {S.~R.}\ \bibnamefont
  {Stroberg}},\ and\ \bibinfo {author} {\bibfnamefont {I.}~\bibnamefont
  {Vernon}},\ }\href {https://doi.org/10.1038/s41567-022-01715-8} {\bibfield
  {journal} {\bibinfo  {journal} {Nat. Phys.}\ }\textbf {\bibinfo {volume}
  {18}},\ \bibinfo {pages} {1196} (\bibinfo {year} {2022})}\BibitemShut
  {NoStop}%
\bibitem [{\citenamefont {Sakurai}(2010)}]{RIBF}%
  \BibitemOpen
  \bibfield  {author} {\bibinfo {author} {\bibfnamefont {H.}~\bibnamefont
  {Sakurai}},\ }\href {https://doi.org/10.1007/s12043-010-0124-6} {\bibfield
  {journal} {\bibinfo  {journal} {Pramana}\ }\textbf {\bibinfo {volume} {75}},\
  \bibinfo {pages} {369} (\bibinfo {year} {2010})}\BibitemShut {NoStop}%
\bibitem [{\citenamefont {Gade}\ and\ \citenamefont {Sherrill}(2016)}]{FRIB}%
  \BibitemOpen
  \bibfield  {author} {\bibinfo {author} {\bibfnamefont {A.}~\bibnamefont
  {Gade}}\ and\ \bibinfo {author} {\bibfnamefont {B.~M.}\ \bibnamefont
  {Sherrill}},\ }\href {https://doi.org/10.1088/0031-8949/91/5/053003}
  {\bibfield  {journal} {\bibinfo  {journal} {Phys. Scr.}\ }\textbf {\bibinfo
  {volume} {91}},\ \bibinfo {pages} {053003} (\bibinfo {year}
  {2016})}\BibitemShut {NoStop}%
\bibitem [{\citenamefont {Aumann}(2007)}]{FAIR}%
  \BibitemOpen
  \bibfield  {author} {\bibinfo {author} {\bibfnamefont {T.}~\bibnamefont
  {Aumann}},\ }\href
  {https://doi.org/https://doi.org/10.1016/j.ppnp.2006.12.018} {\bibfield
  {journal} {\bibinfo  {journal} {Prog. Part. Nucl. Phys.}\ }\textbf {\bibinfo
  {volume} {59}},\ \bibinfo {pages} {3} (\bibinfo {year} {2007})}\BibitemShut
  {NoStop}%
\bibitem [{\citenamefont {Borge}(2016)}]{ISOLDE}%
  \BibitemOpen
  \bibfield  {author} {\bibinfo {author} {\bibfnamefont {M.}~\bibnamefont
  {Borge}},\ }\href
  {https://doi.org/https://doi.org/10.1016/j.nimb.2015.12.048} {\bibfield
  {journal} {\bibinfo  {journal} {Nucl. Instrum. Methods Phys. Res., Sect. B}\
  }\textbf {\bibinfo {volume} {376}},\ \bibinfo {pages} {408} (\bibinfo {year}
  {2016})}\BibitemShut {NoStop}%
\bibitem [{\citenamefont {Gales}(2011)}]{GANIL}%
  \BibitemOpen
  \bibfield  {author} {\bibinfo {author} {\bibfnamefont {S.}~\bibnamefont
  {Gales}},\ }\href {https://doi.org/10.1088/1742-6596/267/1/012009} {\bibfield
   {journal} {\bibinfo  {journal} {J. Phys. Conf. Ser.}\ }\textbf {\bibinfo
  {volume} {267}},\ \bibinfo {pages} {012009} (\bibinfo {year}
  {2011})}\BibitemShut {NoStop}%
\bibitem [{\citenamefont {Dilling}\ \emph {et~al.}(2014)\citenamefont
  {Dilling}, \citenamefont {Krücken},\ and\ \citenamefont {Merminga}}]{ARIEL}%
  \BibitemOpen
  \bibfield  {author} {\bibinfo {author} {\bibfnamefont {J.}~\bibnamefont
  {Dilling}}, \bibinfo {author} {\bibfnamefont {R.}~\bibnamefont {Krücken}},\
  and\ \bibinfo {author} {\bibfnamefont {L.}~\bibnamefont {Merminga}},\ }\href
  {https://doi.org/https://doi.org/10.1007/978-94-007-7963-1} {\emph {\bibinfo
  {title} {ISAC and ARIEL: The TRIUMF Radioactive Beam Facilities and the
  Scientific Program}}},\ \bibinfo {edition} {1st}\ ed.\ (\bibinfo  {publisher}
  {Springer Dordrecht},\ \bibinfo {year} {2014})\BibitemShut {NoStop}%
\bibitem [{\citenamefont {Andrighetto}\ \emph {et~al.}(2018)\citenamefont
  {Andrighetto}, \citenamefont {Manzolaro}, \citenamefont {Corradetti},
  \citenamefont {Scarpa}, \citenamefont {Monetti}, \citenamefont {Rossignoli},
  \citenamefont {Ballan}, \citenamefont {Borgna}, \citenamefont {D’Agostini},
  \citenamefont {Gramegna}, \citenamefont {Prete}, \citenamefont {Meneghetti},
  \citenamefont {Ferrari},\ and\ \citenamefont {Zenoni}}]{SPES}%
  \BibitemOpen
  \bibfield  {author} {\bibinfo {author} {\bibfnamefont {A.}~\bibnamefont
  {Andrighetto}}, \bibinfo {author} {\bibfnamefont {M.}~\bibnamefont
  {Manzolaro}}, \bibinfo {author} {\bibfnamefont {S.}~\bibnamefont
  {Corradetti}}, \bibinfo {author} {\bibfnamefont {D.}~\bibnamefont {Scarpa}},
  \bibinfo {author} {\bibfnamefont {A.}~\bibnamefont {Monetti}}, \bibinfo
  {author} {\bibfnamefont {M.}~\bibnamefont {Rossignoli}}, \bibinfo {author}
  {\bibfnamefont {M.}~\bibnamefont {Ballan}}, \bibinfo {author} {\bibfnamefont
  {F.}~\bibnamefont {Borgna}}, \bibinfo {author} {\bibfnamefont
  {F.}~\bibnamefont {D’Agostini}}, \bibinfo {author} {\bibfnamefont
  {F.}~\bibnamefont {Gramegna}}, \bibinfo {author} {\bibfnamefont
  {G.}~\bibnamefont {Prete}}, \bibinfo {author} {\bibfnamefont
  {G.}~\bibnamefont {Meneghetti}}, \bibinfo {author} {\bibfnamefont
  {M.}~\bibnamefont {Ferrari}},\ and\ \bibinfo {author} {\bibfnamefont
  {A.}~\bibnamefont {Zenoni}},\ }\href
  {https://doi.org/10.1088/1742-6596/966/1/012028} {\bibfield  {journal}
  {\bibinfo  {journal} {J. Phys. Conf. Ser.}\ }\textbf {\bibinfo {volume}
  {966}},\ \bibinfo {pages} {012028} (\bibinfo {year} {2018})}\BibitemShut
  {NoStop}%
\bibitem [{\citenamefont {Otsuka}\ \emph {et~al.}(2020)\citenamefont {Otsuka},
  \citenamefont {Gade}, \citenamefont {Sorlin}, \citenamefont {Suzuki},\ and\
  \citenamefont {Utsuno}}]{Otsuka2021}%
  \BibitemOpen
  \bibfield  {author} {\bibinfo {author} {\bibfnamefont {T.}~\bibnamefont
  {Otsuka}}, \bibinfo {author} {\bibfnamefont {A.}~\bibnamefont {Gade}},
  \bibinfo {author} {\bibfnamefont {O.}~\bibnamefont {Sorlin}}, \bibinfo
  {author} {\bibfnamefont {T.}~\bibnamefont {Suzuki}},\ and\ \bibinfo {author}
  {\bibfnamefont {Y.}~\bibnamefont {Utsuno}},\ }\href
  {https://doi.org/10.1103/RevModPhys.92.015002} {\bibfield  {journal}
  {\bibinfo  {journal} {Rev. Mod. Phys.}\ }\textbf {\bibinfo {volume} {92}},\
  \bibinfo {pages} {015002} (\bibinfo {year} {2020})}\BibitemShut {NoStop}%
\bibitem [{\citenamefont {Tsunoda}\ \emph {et~al.}(2020)\citenamefont
  {Tsunoda}, \citenamefont {Otsuka}, \citenamefont {Takayanagi}, \citenamefont
  {Shimizu}, \citenamefont {Suzuki}, \citenamefont {Utsuno}, \citenamefont
  {Yoshida},\ and\ \citenamefont {Ueno}}]{Tsunoda2020}%
  \BibitemOpen
  \bibfield  {author} {\bibinfo {author} {\bibfnamefont {N.}~\bibnamefont
  {Tsunoda}}, \bibinfo {author} {\bibfnamefont {T.}~\bibnamefont {Otsuka}},
  \bibinfo {author} {\bibfnamefont {K.}~\bibnamefont {Takayanagi}}, \bibinfo
  {author} {\bibfnamefont {N.}~\bibnamefont {Shimizu}}, \bibinfo {author}
  {\bibfnamefont {T.}~\bibnamefont {Suzuki}}, \bibinfo {author} {\bibfnamefont
  {Y.}~\bibnamefont {Utsuno}}, \bibinfo {author} {\bibfnamefont
  {S.}~\bibnamefont {Yoshida}},\ and\ \bibinfo {author} {\bibfnamefont
  {H.}~\bibnamefont {Ueno}},\ }\href
  {https://www.nature.com/articles/s41586-020-2848-x} {\bibfield  {journal}
  {\bibinfo  {journal} {Nature}\ }\textbf {\bibinfo {volume} {587}},\ \bibinfo
  {pages} {66} (\bibinfo {year} {2020})}\BibitemShut {NoStop}%
\bibitem [{\citenamefont {Ravlic}\ \emph {et~al.}(2023)\citenamefont {Ravlic},
  \citenamefont {Y{\"u}ksel}, \citenamefont {Niksic},\ and\ \citenamefont
  {Paar}}]{Ravlic2023}%
  \BibitemOpen
  \bibfield  {author} {\bibinfo {author} {\bibfnamefont {A.}~\bibnamefont
  {Ravlic}}, \bibinfo {author} {\bibfnamefont {E.}~\bibnamefont {Y{\"u}ksel}},
  \bibinfo {author} {\bibfnamefont {T.}~\bibnamefont {Niksic}},\ and\ \bibinfo
  {author} {\bibfnamefont {N.}~\bibnamefont {Paar}},\ }\href
  {https://doi.org/10.1038/s41467-023-40613-2} {\bibfield  {journal} {\bibinfo
  {journal} {Nat. Commun.}\ }\textbf {\bibinfo {volume} {14}},\ \bibinfo
  {pages} {4834} (\bibinfo {year} {2023})}\BibitemShut {NoStop}%
\bibitem [{\citenamefont {Cowan}\ \emph {et~al.}(2021)\citenamefont {Cowan},
  \citenamefont {Sneden}, \citenamefont {Lawler}, \citenamefont {Aprahamian},
  \citenamefont {Wiescher}, \citenamefont {Langanke}, \citenamefont
  {Mart\'{\i}nez-Pinedo},\ and\ \citenamefont {Thielemann}}]{Cowan2021}%
  \BibitemOpen
  \bibfield  {author} {\bibinfo {author} {\bibfnamefont {J.~J.}\ \bibnamefont
  {Cowan}}, \bibinfo {author} {\bibfnamefont {C.}~\bibnamefont {Sneden}},
  \bibinfo {author} {\bibfnamefont {J.~E.}\ \bibnamefont {Lawler}}, \bibinfo
  {author} {\bibfnamefont {A.}~\bibnamefont {Aprahamian}}, \bibinfo {author}
  {\bibfnamefont {M.}~\bibnamefont {Wiescher}}, \bibinfo {author}
  {\bibfnamefont {K.}~\bibnamefont {Langanke}}, \bibinfo {author}
  {\bibfnamefont {G.}~\bibnamefont {Mart\'{\i}nez-Pinedo}},\ and\ \bibinfo
  {author} {\bibfnamefont {F.-K.}\ \bibnamefont {Thielemann}},\ }\href
  {https://doi.org/10.1103/RevModPhys.93.015002} {\bibfield  {journal}
  {\bibinfo  {journal} {Rev. Mod. Phys.}\ }\textbf {\bibinfo {volume} {93}},\
  \bibinfo {pages} {015002} (\bibinfo {year} {2021})}\BibitemShut {NoStop}%
\bibitem [{\citenamefont {Johnson}(2019)}]{Johnson2019}%
  \BibitemOpen
  \bibfield  {author} {\bibinfo {author} {\bibfnamefont {J.~A.}\ \bibnamefont
  {Johnson}},\ }\href {https://doi.org/10.1126/science.aau9540} {\bibfield
  {journal} {\bibinfo  {journal} {Science}\ }\textbf {\bibinfo {volume}
  {363}},\ \bibinfo {pages} {474} (\bibinfo {year} {2019})}\BibitemShut
  {NoStop}%
\bibitem [{\citenamefont {Baroni}\ \emph {et~al.}(2013)\citenamefont {Baroni},
  \citenamefont {Navr\'atil},\ and\ \citenamefont {Quaglioni}}]{Baroni2013}%
  \BibitemOpen
  \bibfield  {author} {\bibinfo {author} {\bibfnamefont {S.}~\bibnamefont
  {Baroni}}, \bibinfo {author} {\bibfnamefont {P.}~\bibnamefont {Navr\'atil}},\
  and\ \bibinfo {author} {\bibfnamefont {S.}~\bibnamefont {Quaglioni}},\ }\href
  {https://doi.org/10.1103/PhysRevLett.110.022505} {\bibfield  {journal}
  {\bibinfo  {journal} {Phys. Rev. Lett.}\ }\textbf {\bibinfo {volume} {110}},\
  \bibinfo {pages} {022505} (\bibinfo {year} {2013})}\BibitemShut {NoStop}%
\bibitem [{\citenamefont {Rotureau}\ \emph {et~al.}(2018)\citenamefont
  {Rotureau}, \citenamefont {Danielewicz}, \citenamefont {Hagen}, \citenamefont
  {Jansen},\ and\ \citenamefont {Nunes}}]{Rotureau2018}%
  \BibitemOpen
  \bibfield  {author} {\bibinfo {author} {\bibfnamefont {J.}~\bibnamefont
  {Rotureau}}, \bibinfo {author} {\bibfnamefont {P.}~\bibnamefont
  {Danielewicz}}, \bibinfo {author} {\bibfnamefont {G.}~\bibnamefont {Hagen}},
  \bibinfo {author} {\bibfnamefont {G.~R.}\ \bibnamefont {Jansen}},\ and\
  \bibinfo {author} {\bibfnamefont {F.~M.}\ \bibnamefont {Nunes}},\ }\href
  {https://doi.org/10.1103/PhysRevC.98.044625} {\bibfield  {journal} {\bibinfo
  {journal} {Phys. Rev. C}\ }\textbf {\bibinfo {volume} {98}},\ \bibinfo
  {pages} {044625} (\bibinfo {year} {2018})}\BibitemShut {NoStop}%
\bibitem [{\citenamefont {Idini}\ \emph {et~al.}(2019)\citenamefont {Idini},
  \citenamefont {Barbieri},\ and\ \citenamefont {Navr\'atil}}]{Idini2019}%
  \BibitemOpen
  \bibfield  {author} {\bibinfo {author} {\bibfnamefont {A.}~\bibnamefont
  {Idini}}, \bibinfo {author} {\bibfnamefont {C.}~\bibnamefont {Barbieri}},\
  and\ \bibinfo {author} {\bibfnamefont {P.}~\bibnamefont {Navr\'atil}},\
  }\href {https://doi.org/10.1103/PhysRevLett.123.092501} {\bibfield  {journal}
  {\bibinfo  {journal} {Phys. Rev. Lett.}\ }\textbf {\bibinfo {volume} {123}},\
  \bibinfo {pages} {092501} (\bibinfo {year} {2019})}\BibitemShut {NoStop}%
\bibitem [{\citenamefont {Burrows}\ \emph {et~al.}(2024)\citenamefont
  {Burrows}, \citenamefont {Launey}, \citenamefont {Mercenne}, \citenamefont
  {Baker}, \citenamefont {Sargsyan}, \citenamefont {Dytrych},\ and\
  \citenamefont {Langr}}]{Burrows2024}%
  \BibitemOpen
  \bibfield  {author} {\bibinfo {author} {\bibfnamefont {M.}~\bibnamefont
  {Burrows}}, \bibinfo {author} {\bibfnamefont {K.~D.}\ \bibnamefont {Launey}},
  \bibinfo {author} {\bibfnamefont {A.}~\bibnamefont {Mercenne}}, \bibinfo
  {author} {\bibfnamefont {R.~B.}\ \bibnamefont {Baker}}, \bibinfo {author}
  {\bibfnamefont {G.~H.}\ \bibnamefont {Sargsyan}}, \bibinfo {author}
  {\bibfnamefont {T.}~\bibnamefont {Dytrych}},\ and\ \bibinfo {author}
  {\bibfnamefont {D.}~\bibnamefont {Langr}},\ }\href
  {https://doi.org/10.1103/PhysRevC.109.014616} {\bibfield  {journal} {\bibinfo
   {journal} {Phys. Rev. C}\ }\textbf {\bibinfo {volume} {109}},\ \bibinfo
  {pages} {014616} (\bibinfo {year} {2024})}\BibitemShut {NoStop}%
\bibitem [{\citenamefont {Vorabbi}\ \emph {et~al.}(2024)\citenamefont
  {Vorabbi}, \citenamefont {Barbieri}, \citenamefont {Som\`a}, \citenamefont
  {Finelli},\ and\ \citenamefont {Giusti}}]{Vorabbi2024}%
  \BibitemOpen
  \bibfield  {author} {\bibinfo {author} {\bibfnamefont {M.}~\bibnamefont
  {Vorabbi}}, \bibinfo {author} {\bibfnamefont {C.}~\bibnamefont {Barbieri}},
  \bibinfo {author} {\bibfnamefont {V.}~\bibnamefont {Som\`a}}, \bibinfo
  {author} {\bibfnamefont {P.}~\bibnamefont {Finelli}},\ and\ \bibinfo {author}
  {\bibfnamefont {C.}~\bibnamefont {Giusti}},\ }\href
  {https://doi.org/10.1103/PhysRevC.109.034613} {\bibfield  {journal} {\bibinfo
   {journal} {Phys. Rev. C}\ }\textbf {\bibinfo {volume} {109}},\ \bibinfo
  {pages} {034613} (\bibinfo {year} {2024})}\BibitemShut {NoStop}%
\bibitem [{\citenamefont {Capuzzi}\ and\ \citenamefont
  {Mahaux}(1996)}]{Capuzzi1996}%
  \BibitemOpen
  \bibfield  {author} {\bibinfo {author} {\bibfnamefont {F.}~\bibnamefont
  {Capuzzi}}\ and\ \bibinfo {author} {\bibfnamefont {C.}~\bibnamefont
  {Mahaux}},\ }\href {https://doi.org/https://doi.org/10.1006/aphy.1996.0006}
  {\bibfield  {journal} {\bibinfo  {journal} {Ann. Phys. (N.Y.)}\ }\textbf
  {\bibinfo {volume} {245}},\ \bibinfo {pages} {147} (\bibinfo {year}
  {1996})}\BibitemShut {NoStop}%
\bibitem [{\citenamefont {Cederbaum}(2001)}]{Cederbaum2001}%
  \BibitemOpen
  \bibfield  {author} {\bibinfo {author} {\bibfnamefont {L.}~\bibnamefont
  {Cederbaum}},\ }\href
  {https://doi.org/https://doi.org/10.1006/aphy.2001.6157} {\bibfield
  {journal} {\bibinfo  {journal} {Ann. Phys. (N.Y.)}\ }\textbf {\bibinfo
  {volume} {291}},\ \bibinfo {pages} {169} (\bibinfo {year}
  {2001})}\BibitemShut {NoStop}%
\bibitem [{\citenamefont {Escher}\ and\ \citenamefont
  {Jennings}(2002)}]{Escher2002}%
  \BibitemOpen
  \bibfield  {author} {\bibinfo {author} {\bibfnamefont {J.}~\bibnamefont
  {Escher}}\ and\ \bibinfo {author} {\bibfnamefont {B.~K.}\ \bibnamefont
  {Jennings}},\ }\href {https://doi.org/10.1103/PhysRevC.66.034313} {\bibfield
  {journal} {\bibinfo  {journal} {Phys. Rev. C}\ }\textbf {\bibinfo {volume}
  {66}},\ \bibinfo {pages} {034313} (\bibinfo {year} {2002})}\BibitemShut
  {NoStop}%
\bibitem [{\citenamefont {Dickhoff}\ and\ \citenamefont
  {Van~Neck}(2005)}]{Dickhoff:manybody}%
  \BibitemOpen
  \bibfield  {author} {\bibinfo {author} {\bibfnamefont {W.}~\bibnamefont
  {Dickhoff}}\ and\ \bibinfo {author} {\bibfnamefont {D.}~\bibnamefont
  {Van~Neck}},\ }\href {https://doi.org/https://doi.org/10.1142/6821} {\emph
  {\bibinfo {title} {Many-Body Theory Exposed!: Propagator Description of
  Quantum Mechanics in Many-Body Systems}}}\ (\bibinfo  {publisher} {World
  Scientific},\ \bibinfo {address} {Singapore},\ \bibinfo {year}
  {2005})\BibitemShut {NoStop}%
\bibitem [{\citenamefont {Mahaux}(1995)}]{Mahaux1995}%
  \BibitemOpen
  \bibfield  {author} {\bibinfo {author} {\bibfnamefont {C.}~\bibnamefont
  {Mahaux}},\ }\bibinfo {title} {Microscopic theories of atomic and nuclear
  optical potentials},\ in\ \href
  {https://doi.org/10.1007/978-1-4615-1937-9_17} {\emph {\bibinfo {booktitle}
  {Recent Progress in Many-Body Theories: Volume 4}}}\ (\bibinfo  {publisher}
  {Springer US},\ \bibinfo {address} {Boston, MA},\ \bibinfo {year} {1995})\
  pp.\ \bibinfo {pages} {171--175}\BibitemShut {NoStop}%
\bibitem [{\citenamefont {Barbieri}\ and\ \citenamefont
  {Carbone}(2017)}]{Barbieri::LectNotesPhys936}%
  \BibitemOpen
  \bibfield  {author} {\bibinfo {author} {\bibfnamefont {C.}~\bibnamefont
  {Barbieri}}\ and\ \bibinfo {author} {\bibfnamefont {A.}~\bibnamefont
  {Carbone}},\ }\href {https://doi.org/10.1007/978-3-319-53336-0_11} {\emph
  {\bibinfo {title} {{Self-Consistent Green's Function Approaches}}}},\ Vol.\
  \bibinfo {volume} {{936 of Lecture Notes in Physics}}\ (\bibinfo  {publisher}
  {Springer},\ \bibinfo {address} {New York},\ \bibinfo {year} {2017})\
  Chap.~\bibinfo {chapter} {11}, pp.\ \bibinfo {pages} {571--644}\BibitemShut
  {NoStop}%
\bibitem [{\citenamefont {Waldecker}\ \emph {et~al.}(2011)\citenamefont
  {Waldecker}, \citenamefont {Barbieri},\ and\ \citenamefont
  {Dickhoff}}]{Waldecker2011}%
  \BibitemOpen
  \bibfield  {author} {\bibinfo {author} {\bibfnamefont {S.~J.}\ \bibnamefont
  {Waldecker}}, \bibinfo {author} {\bibfnamefont {C.}~\bibnamefont
  {Barbieri}},\ and\ \bibinfo {author} {\bibfnamefont {W.~H.}\ \bibnamefont
  {Dickhoff}},\ }\href {https://doi.org/10.1103/PhysRevC.84.034616} {\bibfield
  {journal} {\bibinfo  {journal} {Phys. Rev. C}\ }\textbf {\bibinfo {volume}
  {84}},\ \bibinfo {pages} {034616} (\bibinfo {year} {2011})}\BibitemShut
  {NoStop}%
\bibitem [{\citenamefont {Raimondi}\ and\ \citenamefont
  {Barbieri}(2018)}]{Raimondi2018}%
  \BibitemOpen
  \bibfield  {author} {\bibinfo {author} {\bibfnamefont {F.}~\bibnamefont
  {Raimondi}}\ and\ \bibinfo {author} {\bibfnamefont {C.}~\bibnamefont
  {Barbieri}},\ }\href {https://doi.org/10.1103/PhysRevC.97.054308} {\bibfield
  {journal} {\bibinfo  {journal} {Phys. Rev. C}\ }\textbf {\bibinfo {volume}
  {97}},\ \bibinfo {pages} {054308} (\bibinfo {year} {2018})}\BibitemShut
  {NoStop}%
\bibitem [{\citenamefont {Prokof'ev}\ and\ \citenamefont
  {Svistunov}(1998)}]{Prokofev1998}%
  \BibitemOpen
  \bibfield  {author} {\bibinfo {author} {\bibfnamefont {N.~V.}\ \bibnamefont
  {Prokof'ev}}\ and\ \bibinfo {author} {\bibfnamefont {B.~V.}\ \bibnamefont
  {Svistunov}},\ }\href {https://doi.org/10.1103/PhysRevLett.81.2514}
  {\bibfield  {journal} {\bibinfo  {journal} {Phys. Rev. Lett.}\ }\textbf
  {\bibinfo {volume} {81}},\ \bibinfo {pages} {2514} (\bibinfo {year}
  {1998})}\BibitemShut {NoStop}%
\bibitem [{\citenamefont {Prokof'ev}\ and\ \citenamefont
  {Svistunov}(2008)}]{Prokofev2008}%
  \BibitemOpen
  \bibfield  {author} {\bibinfo {author} {\bibfnamefont {N.~V.}\ \bibnamefont
  {Prokof'ev}}\ and\ \bibinfo {author} {\bibfnamefont {B.~V.}\ \bibnamefont
  {Svistunov}},\ }\href {https://doi.org/10.1103/PhysRevB.77.125101} {\bibfield
   {journal} {\bibinfo  {journal} {Phys. Rev. B}\ }\textbf {\bibinfo {volume}
  {77}},\ \bibinfo {pages} {125101} (\bibinfo {year} {2008})}\BibitemShut
  {NoStop}%
\bibitem [{\citenamefont {Van~Houcke}\ \emph {et~al.}(2019)\citenamefont
  {Van~Houcke}, \citenamefont {Werner}, \citenamefont {Ohgoe}, \citenamefont
  {Prokof'ev},\ and\ \citenamefont {Svistunov}}]{VanHoucke2019}%
  \BibitemOpen
  \bibfield  {author} {\bibinfo {author} {\bibfnamefont {K.}~\bibnamefont
  {Van~Houcke}}, \bibinfo {author} {\bibfnamefont {F.}~\bibnamefont {Werner}},
  \bibinfo {author} {\bibfnamefont {T.}~\bibnamefont {Ohgoe}}, \bibinfo
  {author} {\bibfnamefont {N.~V.}\ \bibnamefont {Prokof'ev}},\ and\ \bibinfo
  {author} {\bibfnamefont {B.~V.}\ \bibnamefont {Svistunov}},\ }\href
  {https://doi.org/10.1103/PhysRevB.99.035140} {\bibfield  {journal} {\bibinfo
  {journal} {Phys. Rev. B}\ }\textbf {\bibinfo {volume} {99}},\ \bibinfo
  {pages} {035140} (\bibinfo {year} {2019})}\BibitemShut {NoStop}%
\bibitem [{\citenamefont {Van~Houcke}\ \emph {et~al.}(2012)\citenamefont
  {Van~Houcke}, \citenamefont {Werner}, \citenamefont {Kozik}, \citenamefont
  {Prokof'ev}, \citenamefont {Svistunov}, \citenamefont {Ku}, \citenamefont
  {Sommer}, \citenamefont {Cheuk}, \citenamefont {Schirotzek},\ and\
  \citenamefont {Zwierlein}}]{VanHoucke2012}%
  \BibitemOpen
  \bibfield  {author} {\bibinfo {author} {\bibfnamefont {K.}~\bibnamefont
  {Van~Houcke}}, \bibinfo {author} {\bibfnamefont {F.}~\bibnamefont {Werner}},
  \bibinfo {author} {\bibfnamefont {E.}~\bibnamefont {Kozik}}, \bibinfo
  {author} {\bibfnamefont {N.}~\bibnamefont {Prokof'ev}}, \bibinfo {author}
  {\bibfnamefont {B.}~\bibnamefont {Svistunov}}, \bibinfo {author}
  {\bibfnamefont {M.~J.~H.}\ \bibnamefont {Ku}}, \bibinfo {author}
  {\bibfnamefont {A.~T.}\ \bibnamefont {Sommer}}, \bibinfo {author}
  {\bibfnamefont {L.~W.}\ \bibnamefont {Cheuk}}, \bibinfo {author}
  {\bibfnamefont {A.}~\bibnamefont {Schirotzek}},\ and\ \bibinfo {author}
  {\bibfnamefont {M.~W.}\ \bibnamefont {Zwierlein}},\ }\href
  {https://doi.org/10.1038/nphys2273} {\bibfield  {journal} {\bibinfo
  {journal} {Nat. Phys.}\ }\textbf {\bibinfo {volume} {8}},\ \bibinfo {pages}
  {366} (\bibinfo {year} {2012})}\BibitemShut {NoStop}%
\bibitem [{\citenamefont {Haule}\ and\ \citenamefont {Chen}(2022)}]{Haule2022}%
  \BibitemOpen
  \bibfield  {author} {\bibinfo {author} {\bibfnamefont {K.}~\bibnamefont
  {Haule}}\ and\ \bibinfo {author} {\bibfnamefont {K.}~\bibnamefont {Chen}},\
  }\href {https://doi.org/10.1038/s41598-022-06188-6} {\bibfield  {journal}
  {\bibinfo  {journal} {Sci. Rep.}\ }\textbf {\bibinfo {volume} {12}},\
  \bibinfo {pages} {2294} (\bibinfo {year} {2022})}\BibitemShut {NoStop}%
\bibitem [{\citenamefont {Chen}\ and\ \citenamefont {Haule}(2019)}]{Chen2019}%
  \BibitemOpen
  \bibfield  {author} {\bibinfo {author} {\bibfnamefont {K.}~\bibnamefont
  {Chen}}\ and\ \bibinfo {author} {\bibfnamefont {K.}~\bibnamefont {Haule}},\
  }\href {https://doi.org/10.1038/s41467-019-11708-6} {\bibfield  {journal}
  {\bibinfo  {journal} {Nat. Commun.}\ }\textbf {\bibinfo {volume} {10}},\
  \bibinfo {pages} {3725} (\bibinfo {year} {2019})}\BibitemShut {NoStop}%
\bibitem [{\citenamefont {Vanhoecke}\ and\ \citenamefont
  {Schir{\`o}}(2024)}]{Vanhoecke2024}%
  \BibitemOpen
  \bibfield  {author} {\bibinfo {author} {\bibfnamefont {M.}~\bibnamefont
  {Vanhoecke}}\ and\ \bibinfo {author} {\bibfnamefont {M.}~\bibnamefont
  {Schir{\`o}}},\ }\href {https://doi.org/10.1103/PhysRevB.109.125125}
  {\bibfield  {journal} {\bibinfo  {journal} {Phys. Rev. B.}\ }\textbf
  {\bibinfo {volume} {109}},\ \bibinfo {pages} {125125} (\bibinfo {year}
  {2024})}\BibitemShut {NoStop}%
\bibitem [{\citenamefont {Baerdemacker}\ \emph {et~al.}(2014)\citenamefont
  {Baerdemacker}, \citenamefont {Hellemans}, \citenamefont {van~den Berg},
  \citenamefont {Caux}, \citenamefont {Heyde}, \citenamefont {Raemdonck},
  \citenamefont {Neck},\ and\ \citenamefont {Johnson}}]{Baerdemacker2014Sn}%
  \BibitemOpen
  \bibfield  {author} {\bibinfo {author} {\bibfnamefont {S.~D.}\ \bibnamefont
  {Baerdemacker}}, \bibinfo {author} {\bibfnamefont {V.}~\bibnamefont
  {Hellemans}}, \bibinfo {author} {\bibfnamefont {R.}~\bibnamefont {van~den
  Berg}}, \bibinfo {author} {\bibfnamefont {J.-S.}\ \bibnamefont {Caux}},
  \bibinfo {author} {\bibfnamefont {K.}~\bibnamefont {Heyde}}, \bibinfo
  {author} {\bibfnamefont {M.~V.}\ \bibnamefont {Raemdonck}}, \bibinfo {author}
  {\bibfnamefont {D.~V.}\ \bibnamefont {Neck}},\ and\ \bibinfo {author}
  {\bibfnamefont {P.~A.}\ \bibnamefont {Johnson}},\ }\href
  {https://doi.org/10.1088/1742-6596/533/1/012058} {\bibfield  {journal}
  {\bibinfo  {journal} {J. Phys. Conf. Ser.}\ }\textbf {\bibinfo {volume}
  {533}},\ \bibinfo {pages} {012058} (\bibinfo {year} {2014})}\BibitemShut
  {NoStop}%
\bibitem [{\citenamefont
  {Richardson}(1963{\natexlab{a}})}]{Richardson:EvenLead}%
  \BibitemOpen
  \bibfield  {author} {\bibinfo {author} {\bibfnamefont {R.~W.}\ \bibnamefont
  {Richardson}},\ }\href
  {https://doi.org/https://doi.org/10.1016/S0375-9601(63)80039-0} {\bibfield
  {journal} {\bibinfo  {journal} {Phys. Lett.}\ }\textbf {\bibinfo {volume}
  {5}},\ \bibinfo {pages} {82} (\bibinfo {year}
  {1963}{\natexlab{a}})}\BibitemShut {NoStop}%
\bibitem [{\citenamefont {Johnson}\ \emph {et~al.}(2020)\citenamefont
  {Johnson}, \citenamefont {Fecteau}, \citenamefont {Berthiaume}, \citenamefont
  {Cloutier}, \citenamefont {Carrier}, \citenamefont {Gratton}, \citenamefont
  {Bultinck}, \citenamefont {De~Baerdemacker}, \citenamefont {Van~Neck},
  \citenamefont {Limacher},\ and\ \citenamefont {Ayers}}]{Johnson2020}%
  \BibitemOpen
  \bibfield  {author} {\bibinfo {author} {\bibfnamefont {P.~A.}\ \bibnamefont
  {Johnson}}, \bibinfo {author} {\bibfnamefont {C.-{\'{E}}.}\ \bibnamefont
  {Fecteau}}, \bibinfo {author} {\bibfnamefont {F.}~\bibnamefont {Berthiaume}},
  \bibinfo {author} {\bibfnamefont {S.}~\bibnamefont {Cloutier}}, \bibinfo
  {author} {\bibfnamefont {L.}~\bibnamefont {Carrier}}, \bibinfo {author}
  {\bibfnamefont {M.}~\bibnamefont {Gratton}}, \bibinfo {author} {\bibfnamefont
  {P.}~\bibnamefont {Bultinck}}, \bibinfo {author} {\bibfnamefont
  {S.}~\bibnamefont {De~Baerdemacker}}, \bibinfo {author} {\bibfnamefont
  {D.}~\bibnamefont {Van~Neck}}, \bibinfo {author} {\bibfnamefont
  {P.}~\bibnamefont {Limacher}},\ and\ \bibinfo {author} {\bibfnamefont
  {P.~W.}\ \bibnamefont {Ayers}},\ }\href {https://doi.org/10.1063/5.0022189}
  {\bibfield  {journal} {\bibinfo  {journal} {J. Chem. Phys.}\ }\textbf
  {\bibinfo {volume} {153}},\ \bibinfo {pages} {104110} (\bibinfo {year}
  {2020})}\BibitemShut {NoStop}%
\bibitem [{\citenamefont {Hjorth-Jensen}\ \emph {et~al.}(2017)\citenamefont
  {Hjorth-Jensen}, \citenamefont {Lombardo},\ and\ \citenamefont {van
  Kolck}}]{Hjorth-Jensen::LectNotesPhys936}%
  \BibitemOpen
  \bibinfo {editor} {\bibfnamefont {M.}~\bibnamefont {Hjorth-Jensen}}, \bibinfo
  {editor} {\bibfnamefont {M.~P.}\ \bibnamefont {Lombardo}},\ and\ \bibinfo
  {editor} {\bibfnamefont {U.}~\bibnamefont {van Kolck}},\ eds.,\ \href
  {https://doi.org/10.1007/978-3-319-53336-0} {\emph {\bibinfo {title} {{An
  Advanced Course in Computational Nuclear Physics}}}},\ \bibinfo {series}
  {Lecture Notes in Physics}, Vol.\ \bibinfo {volume} {936}\ (\bibinfo
  {publisher} {Springer},\ \bibinfo {year} {2017})\BibitemShut {NoStop}%
\bibitem [{\citenamefont {Rigo}\ \emph {et~al.}(2023)\citenamefont {Rigo},
  \citenamefont {Hall}, \citenamefont {Hjorth-Jensen}, \citenamefont {Lovato},\
  and\ \citenamefont {Pederiva}}]{Rigo2023}%
  \BibitemOpen
  \bibfield  {author} {\bibinfo {author} {\bibfnamefont {M.}~\bibnamefont
  {Rigo}}, \bibinfo {author} {\bibfnamefont {B.}~\bibnamefont {Hall}}, \bibinfo
  {author} {\bibfnamefont {M.}~\bibnamefont {Hjorth-Jensen}}, \bibinfo {author}
  {\bibfnamefont {A.}~\bibnamefont {Lovato}},\ and\ \bibinfo {author}
  {\bibfnamefont {F.}~\bibnamefont {Pederiva}},\ }\href
  {https://doi.org/10.1103/PhysRevE.107.025310} {\bibfield  {journal} {\bibinfo
   {journal} {Phys. Rev. E}\ }\textbf {\bibinfo {volume} {107}},\ \bibinfo
  {pages} {025310} (\bibinfo {year} {2023})}\BibitemShut {NoStop}%
\bibitem [{\citenamefont {Companys~Franzke}\ \emph {et~al.}(2024)\citenamefont
  {Companys~Franzke}, \citenamefont {Tichai}, \citenamefont {Hebeler},\ and\
  \citenamefont {Schwenk}}]{Franzke2024}%
  \BibitemOpen
  \bibfield  {author} {\bibinfo {author} {\bibfnamefont {M.}~\bibnamefont
  {Companys~Franzke}}, \bibinfo {author} {\bibfnamefont {A.}~\bibnamefont
  {Tichai}}, \bibinfo {author} {\bibfnamefont {K.}~\bibnamefont {Hebeler}},\
  and\ \bibinfo {author} {\bibfnamefont {A.}~\bibnamefont {Schwenk}},\ }\href
  {https://doi.org/10.1103/PhysRevC.109.024311} {\bibfield  {journal} {\bibinfo
   {journal} {Phys. Rev. C}\ }\textbf {\bibinfo {volume} {109}},\ \bibinfo
  {pages} {024311} (\bibinfo {year} {2024})}\BibitemShut {NoStop}%
\bibitem [{\citenamefont
  {Richardson}(1963{\natexlab{b}})}]{Richardson:Eigenstates}%
  \BibitemOpen
  \bibfield  {author} {\bibinfo {author} {\bibfnamefont {R.~W.}\ \bibnamefont
  {Richardson}},\ }\href
  {https://doi.org/https://doi.org/10.1016/0031-9163(63)90259-2} {\bibfield
  {journal} {\bibinfo  {journal} {Phys. Lett.}\ }\textbf {\bibinfo {volume}
  {3}},\ \bibinfo {pages} {277} (\bibinfo {year}
  {1963}{\natexlab{b}})}\BibitemShut {NoStop}%
\bibitem [{Sup()}]{SupplMat}%
  \BibitemOpen
  \href@noop {} {}\bibinfo {howpublished} {See Supplemental Material at
  \url{URL_will_be_inserted_by_publisher} for details about the Richardson
  model, the fundamental equations of DiagMC, the error induced by the finite
  statistics of the Monte Carlo stochastic process, and the convergence of the
  ladder series.}\BibitemShut {Stop}%
\bibitem [{\citenamefont {Brolli}(2023)}]{Brolli::Masters}%
  \BibitemOpen
  \bibfield  {author} {\bibinfo {author} {\bibfnamefont {S.}~\bibnamefont
  {Brolli}},\ }\emph {\bibinfo {title} {A {Diagrammatic Monte Carlo Method for
  a Fermionic Pairing Model}}},\ \href@noop {} {Master's thesis},\ \bibinfo
  {school} {Università degli Studi di Milano} (\bibinfo {year}
  {2023})\BibitemShut {NoStop}%
\bibitem [{\citenamefont {Drischler}\ \emph {et~al.}(2019)\citenamefont
  {Drischler}, \citenamefont {Hebeler},\ and\ \citenamefont
  {Schwenk}}]{Drischler2019}%
  \BibitemOpen
  \bibfield  {author} {\bibinfo {author} {\bibfnamefont {C.}~\bibnamefont
  {Drischler}}, \bibinfo {author} {\bibfnamefont {K.}~\bibnamefont {Hebeler}},\
  and\ \bibinfo {author} {\bibfnamefont {A.}~\bibnamefont {Schwenk}},\ }\href
  {https://doi.org/10.1103/PhysRevLett.122.042501} {\bibfield  {journal}
  {\bibinfo  {journal} {Phys. Rev. Lett.}\ }\textbf {\bibinfo {volume} {122}},\
  \bibinfo {pages} {042501} (\bibinfo {year} {2019})}\BibitemShut {NoStop}%
\bibitem [{\citenamefont {Arthuis}\ \emph {et~al.}(2019)\citenamefont
  {Arthuis}, \citenamefont {Duguet}, \citenamefont {Tichai}, \citenamefont
  {Lasseri},\ and\ \citenamefont {Ebran}}]{ArthuisADG1_2019}%
  \BibitemOpen
  \bibfield  {author} {\bibinfo {author} {\bibfnamefont {P.}~\bibnamefont
  {Arthuis}}, \bibinfo {author} {\bibfnamefont {T.}~\bibnamefont {Duguet}},
  \bibinfo {author} {\bibfnamefont {A.}~\bibnamefont {Tichai}}, \bibinfo
  {author} {\bibfnamefont {R.-D.}\ \bibnamefont {Lasseri}},\ and\ \bibinfo
  {author} {\bibfnamefont {J.-P.}\ \bibnamefont {Ebran}},\ }\href
  {https://doi.org/https://doi.org/10.1016/j.cpc.2018.11.023} {\bibfield
  {journal} {\bibinfo  {journal} {Comput. Phys. Commun.}\ }\textbf {\bibinfo
  {volume} {240}},\ \bibinfo {pages} {202} (\bibinfo {year}
  {2019})}\BibitemShut {NoStop}%
\bibitem [{\citenamefont {Metropolis}\ \emph {et~al.}(1953)\citenamefont
  {Metropolis}, \citenamefont {Rosenbluth}, \citenamefont {Rosenbluth},
  \citenamefont {Teller},\ and\ \citenamefont {Teller}}]{Metropolis1953}%
  \BibitemOpen
  \bibfield  {author} {\bibinfo {author} {\bibfnamefont {N.}~\bibnamefont
  {Metropolis}}, \bibinfo {author} {\bibfnamefont {A.~W.}\ \bibnamefont
  {Rosenbluth}}, \bibinfo {author} {\bibfnamefont {M.~N.}\ \bibnamefont
  {Rosenbluth}}, \bibinfo {author} {\bibfnamefont {A.~H.}\ \bibnamefont
  {Teller}},\ and\ \bibinfo {author} {\bibfnamefont {E.}~\bibnamefont
  {Teller}},\ }\href {https://doi.org/10.1063/1.1699114} {\bibfield  {journal}
  {\bibinfo  {journal} {J. Chem. Phys.}\ }\textbf {\bibinfo {volume} {21}},\
  \bibinfo {pages} {1087} (\bibinfo {year} {1953})}\BibitemShut {NoStop}%
\bibitem [{\citenamefont {Hastings}(1970)}]{Hasting1970}%
  \BibitemOpen
  \bibfield  {author} {\bibinfo {author} {\bibfnamefont {W.~K.}\ \bibnamefont
  {Hastings}},\ }\href {https://doi.org/10.1093/biomet/57.1.97} {\bibfield
  {journal} {\bibinfo  {journal} {Biometrika}\ }\textbf {\bibinfo {volume}
  {57}},\ \bibinfo {pages} {97} (\bibinfo {year} {1970})}\BibitemShut {NoStop}%
\bibitem [{\citenamefont {Schirmer}(2018)}]{Schirmer2018LNC}%
  \BibitemOpen
  \bibfield  {author} {\bibinfo {author} {\bibfnamefont {J.}~\bibnamefont
  {Schirmer}},\ }\href {https://doi.org/10.1007/978-3-319-93602-4} {\emph
  {\bibinfo {title} {Many-Body Methods for Atoms, Molecules and Clusters}}}\
  (\bibinfo  {publisher} {Springer International Publishing},\ \bibinfo
  {address} {Cham},\ \bibinfo {year} {2018})\BibitemShut {NoStop}%
\bibitem [{\citenamefont {Schirmer}\ \emph {et~al.}(1983)\citenamefont
  {Schirmer}, \citenamefont {Cederbaum},\ and\ \citenamefont
  {Walter}}]{Schirmer1983::ADC}%
  \BibitemOpen
  \bibfield  {author} {\bibinfo {author} {\bibfnamefont {J.}~\bibnamefont
  {Schirmer}}, \bibinfo {author} {\bibfnamefont {L.~S.}\ \bibnamefont
  {Cederbaum}},\ and\ \bibinfo {author} {\bibfnamefont {O.}~\bibnamefont
  {Walter}},\ }\href {https://doi.org/10.1103/PhysRevA.28.1237} {\bibfield
  {journal} {\bibinfo  {journal} {Phys. Rev. A}\ }\textbf {\bibinfo {volume}
  {28}},\ \bibinfo {pages} {1237} (\bibinfo {year} {1983})}\BibitemShut
  {NoStop}%
\bibitem [{\citenamefont {Som\`a}\ \emph {et~al.}(2014)\citenamefont {Som\`a},
  \citenamefont {Barbieri},\ and\ \citenamefont {Duguet}}]{Soma2014::Gorkov}%
  \BibitemOpen
  \bibfield  {author} {\bibinfo {author} {\bibfnamefont {V.}~\bibnamefont
  {Som\`a}}, \bibinfo {author} {\bibfnamefont {C.}~\bibnamefont {Barbieri}},\
  and\ \bibinfo {author} {\bibfnamefont {T.}~\bibnamefont {Duguet}},\ }\href
  {https://doi.org/10.1103/PhysRevC.89.024323} {\bibfield  {journal} {\bibinfo
  {journal} {Phys. Rev. C}\ }\textbf {\bibinfo {volume} {89}},\ \bibinfo
  {pages} {024323} (\bibinfo {year} {2014})}\BibitemShut {NoStop}%
\bibitem [{\citenamefont {Barbieri}\ \emph {et~al.}(2022)\citenamefont
  {Barbieri}, \citenamefont {Duguet},\ and\ \citenamefont
  {Som\`a}}]{Barbieri2022::GorkovADC(3)}%
  \BibitemOpen
  \bibfield  {author} {\bibinfo {author} {\bibfnamefont {C.}~\bibnamefont
  {Barbieri}}, \bibinfo {author} {\bibfnamefont {T.}~\bibnamefont {Duguet}},\
  and\ \bibinfo {author} {\bibfnamefont {V.}~\bibnamefont {Som\`a}},\ }\href
  {https://doi.org/10.1103/PhysRevC.105.044330} {\bibfield  {journal} {\bibinfo
   {journal} {Phys. Rev. C}\ }\textbf {\bibinfo {volume} {105}},\ \bibinfo
  {pages} {044330} (\bibinfo {year} {2022})}\BibitemShut {NoStop}%
\bibitem [{\citenamefont {Galitskii}\ and\ \citenamefont
  {Migdal}(1958)}]{Migdal1958}%
  \BibitemOpen
  \bibfield  {author} {\bibinfo {author} {\bibfnamefont {V.~M.}\ \bibnamefont
  {Galitskii}}\ and\ \bibinfo {author} {\bibfnamefont {A.~B.}\ \bibnamefont
  {Migdal}},\ }\href {http://jetp.ras.ru/cgi-bin/e/index/e/7/1/p96?a=list}
  {\bibfield  {journal} {\bibinfo  {journal} {Sov. Phys. JETP}\ }\textbf
  {\bibinfo {volume} {7}},\ \bibinfo {pages} {96} (\bibinfo {year}
  {1958})}\BibitemShut {NoStop}%
\bibitem [{\citenamefont {Koltun}(1974)}]{Koltun1974}%
  \BibitemOpen
  \bibfield  {author} {\bibinfo {author} {\bibfnamefont {D.~S.}\ \bibnamefont
  {Koltun}},\ }\href {https://doi.org/10.1103/PhysRevC.9.484} {\bibfield
  {journal} {\bibinfo  {journal} {Phys. Rev. C}\ }\textbf {\bibinfo {volume}
  {9}},\ \bibinfo {pages} {484} (\bibinfo {year} {1974})}\BibitemShut {NoStop}%
\bibitem [{\citenamefont {Houcke}\ \emph {et~al.}(2010)\citenamefont {Houcke},
  \citenamefont {Kozik}, \citenamefont {Prokof’ev},\ and\ \citenamefont
  {Svistunov}}]{VanHoucke2010}%
  \BibitemOpen
  \bibfield  {author} {\bibinfo {author} {\bibfnamefont {K.~V.}\ \bibnamefont
  {Houcke}}, \bibinfo {author} {\bibfnamefont {E.}~\bibnamefont {Kozik}},
  \bibinfo {author} {\bibfnamefont {N.}~\bibnamefont {Prokof’ev}},\ and\
  \bibinfo {author} {\bibfnamefont {B.}~\bibnamefont {Svistunov}},\ }\href
  {https://doi.org/https://doi.org/10.1016/j.phpro.2010.09.034} {\bibfield
  {journal} {\bibinfo  {journal} {Phys. Procedia}\ }\textbf {\bibinfo {volume}
  {6}},\ \bibinfo {pages} {95} (\bibinfo {year} {2010})}\BibitemShut {NoStop}%
\bibitem [{\citenamefont {Rossi}\ \emph {et~al.}(2018)\citenamefont {Rossi},
  \citenamefont {Ohgoe}, \citenamefont {Van~Houcke},\ and\ \citenamefont
  {Werner}}]{Rossi2018}%
  \BibitemOpen
  \bibfield  {author} {\bibinfo {author} {\bibfnamefont {R.}~\bibnamefont
  {Rossi}}, \bibinfo {author} {\bibfnamefont {T.}~\bibnamefont {Ohgoe}},
  \bibinfo {author} {\bibfnamefont {K.}~\bibnamefont {Van~Houcke}},\ and\
  \bibinfo {author} {\bibfnamefont {F.}~\bibnamefont {Werner}},\ }\href
  {https://doi.org/10.1103/PhysRevLett.121.130405} {\bibfield  {journal}
  {\bibinfo  {journal} {Phys. Rev. Lett.}\ }\textbf {\bibinfo {volume} {121}},\
  \bibinfo {pages} {130405} (\bibinfo {year} {2018})}\BibitemShut {NoStop}%
\end{thebibliography}%


\begin{thebibliography}{6}%
\makeatletter
\providecommand \@ifxundefined [1]{%
 \@ifx{#1\undefined}
}%
\providecommand \@ifnum [1]{%
 \ifnum #1\expandafter \@firstoftwo
 \else \expandafter \@secondoftwo
 \fi
}%
\providecommand \@ifx [1]{%
 \ifx #1\expandafter \@firstoftwo
 \else \expandafter \@secondoftwo
 \fi
}%
\providecommand \natexlab [1]{#1}%
\providecommand \enquote  [1]{``#1''}%
\providecommand \bibnamefont  [1]{#1}%
\providecommand \bibfnamefont [1]{#1}%
\providecommand \citenamefont [1]{#1}%
\providecommand \href@noop [0]{\@secondoftwo}%
\providecommand \href [0]{\begingroup \@sanitize@url \@href}%
\providecommand \@href[1]{\@@startlink{#1}\@@href}%
\providecommand \@@href[1]{\endgroup#1\@@endlink}%
\providecommand \@sanitize@url [0]{\catcode `\\12\catcode `\$12\catcode
  `\&12\catcode `\#12\catcode `\^12\catcode `\_12\catcode `\%12\relax}%
\providecommand \@@startlink[1]{}%
\providecommand \@@endlink[0]{}%
\providecommand \url  [0]{\begingroup\@sanitize@url \@url }%
\providecommand \@url [1]{\endgroup\@href {#1}{\urlprefix }}%
\providecommand \urlprefix  [0]{URL }%
\providecommand \Eprint [0]{\href }%
\providecommand \doibase [0]{https://doi.org/}%
\providecommand \selectlanguage [0]{\@gobble}%
\providecommand \bibinfo  [0]{\@secondoftwo}%
\providecommand \bibfield  [0]{\@secondoftwo}%
\providecommand \translation [1]{[#1]}%
\providecommand \BibitemOpen [0]{}%
\providecommand \bibitemStop [0]{}%
\providecommand \bibitemNoStop [0]{.\EOS\space}%
\providecommand \EOS [0]{\spacefactor3000\relax}%
\providecommand \BibitemShut  [1]{\csname bibitem#1\endcsname}%
\let\auto@bib@innerbib\@empty
\bibitem [{\citenamefont {Richardson}(1963)}]{Richardson:Eigenstates}%
  \BibitemOpen
  \bibfield  {author} {\bibinfo {author} {\bibfnamefont {R.~W.}\ \bibnamefont
  {Richardson}},\ }\href
  {https://doi.org/https://doi.org/10.1016/0031-9163(63)90259-2} {\bibfield
  {journal} {\bibinfo  {journal} {Phys. Lett.}\ }\textbf {\bibinfo {volume}
  {3}},\ \bibinfo {pages} {277} (\bibinfo {year} {1963})}\BibitemShut {NoStop}%
\bibitem [{\citenamefont {Van~Houcke}\ \emph {et~al.}(2019)\citenamefont
  {Van~Houcke}, \citenamefont {Werner}, \citenamefont {Ohgoe}, \citenamefont
  {Prokof'ev},\ and\ \citenamefont {Svistunov}}]{VanHoucke2019}%
  \BibitemOpen
  \bibfield  {author} {\bibinfo {author} {\bibfnamefont {K.}~\bibnamefont
  {Van~Houcke}}, \bibinfo {author} {\bibfnamefont {F.}~\bibnamefont {Werner}},
  \bibinfo {author} {\bibfnamefont {T.}~\bibnamefont {Ohgoe}}, \bibinfo
  {author} {\bibfnamefont {N.~V.}\ \bibnamefont {Prokof'ev}},\ and\ \bibinfo
  {author} {\bibfnamefont {B.~V.}\ \bibnamefont {Svistunov}},\ }\href
  {https://doi.org/10.1103/PhysRevB.99.035140} {\bibfield  {journal} {\bibinfo
  {journal} {Phys. Rev. B}\ }\textbf {\bibinfo {volume} {99}},\ \bibinfo
  {pages} {035140} (\bibinfo {year} {2019})}\BibitemShut {NoStop}%
\bibitem [{\citenamefont {Dickhoff}\ and\ \citenamefont
  {Van~Neck}(2005)}]{Dickhoff:manybody}%
  \BibitemOpen
  \bibfield  {author} {\bibinfo {author} {\bibfnamefont {W.}~\bibnamefont
  {Dickhoff}}\ and\ \bibinfo {author} {\bibfnamefont {D.}~\bibnamefont
  {Van~Neck}},\ }\href {https://doi.org/https://doi.org/10.1142/6821} {\emph
  {\bibinfo {title} {Many-Body Theory Exposed!: Propagator Description of
  Quantum Mechanics in Many-Body Systems}}}\ (\bibinfo  {publisher} {World
  Scientific},\ \bibinfo {address} {Singapore},\ \bibinfo {year}
  {2005})\BibitemShut {NoStop}%
\bibitem [{\citenamefont {Metropolis}\ \emph {et~al.}(1953)\citenamefont
  {Metropolis}, \citenamefont {Rosenbluth}, \citenamefont {Rosenbluth},
  \citenamefont {Teller},\ and\ \citenamefont {Teller}}]{Metropolis1953}%
  \BibitemOpen
  \bibfield  {author} {\bibinfo {author} {\bibfnamefont {N.}~\bibnamefont
  {Metropolis}}, \bibinfo {author} {\bibfnamefont {A.~W.}\ \bibnamefont
  {Rosenbluth}}, \bibinfo {author} {\bibfnamefont {M.~N.}\ \bibnamefont
  {Rosenbluth}}, \bibinfo {author} {\bibfnamefont {A.~H.}\ \bibnamefont
  {Teller}},\ and\ \bibinfo {author} {\bibfnamefont {E.}~\bibnamefont
  {Teller}},\ }\href {https://doi.org/10.1063/1.1699114} {\bibfield  {journal}
  {\bibinfo  {journal} {J. Chem. Phys.}\ }\textbf {\bibinfo {volume} {21}},\
  \bibinfo {pages} {1087} (\bibinfo {year} {1953})}\BibitemShut {NoStop}%
\bibitem [{\citenamefont {Hastings}(1970)}]{Hasting1970}%
  \BibitemOpen
  \bibfield  {author} {\bibinfo {author} {\bibfnamefont {W.~K.}\ \bibnamefont
  {Hastings}},\ }\href {https://doi.org/10.1093/biomet/57.1.97} {\bibfield
  {journal} {\bibinfo  {journal} {Biometrika}\ }\textbf {\bibinfo {volume}
  {57}},\ \bibinfo {pages} {97} (\bibinfo {year} {1970})}\BibitemShut {NoStop}%
\bibitem [{\citenamefont {Barbieri}\ and\ \citenamefont
  {Carbone}(2017)}]{Barbieri::LectNotesPhys936}%
  \BibitemOpen
  \bibfield  {author} {\bibinfo {author} {\bibfnamefont {C.}~\bibnamefont
  {Barbieri}}\ and\ \bibinfo {author} {\bibfnamefont {A.}~\bibnamefont
  {Carbone}},\ }\href {https://doi.org/10.1007/978-3-319-53336-0_11} {\emph
  {\bibinfo {title} {{Self-Consistent Green's Function Approaches}}}},\ Vol.\
  \bibinfo {volume} {{936 of Lecture Notes in Physics}}\ (\bibinfo  {publisher}
  {Springer},\ \bibinfo {address} {New York},\ \bibinfo {year} {2017})\
  Chap.~\bibinfo {chapter} {11}, pp.\ \bibinfo {pages} {571--644}\BibitemShut
  {NoStop}%
\end{thebibliography}%

\end{document}


\begin{center}
\section*{Supplemental Material}
\end{center}

\renewcommand{\thefigure}{S\arabic{figure}}
\renewcommand{\thetable}{S\arabic{table}}
\renewcommand{\theequation}{S\arabic{equation}}
\renewcommand{\thepage}{\arabic{page}}
\setcounter{figure}{0}
\setcounter{table}{0}
\setcounter{equation}{0}
\setcounter{page}{1} 

\subsection*{Exact solution of the Richardson model}

The Richardson Hamiltonian has an exact analytic solution~\cite{Richardson:Eigenstates}. For $P$ pairs of fermions, this is

\begin{equation}
\ket{\Psi_{g.s.}^{2P}} = \prod_{i = 1}^{P} b_{i}^{\dagger} \ket{0},
\end{equation}

\noindent where

\begin{equation}
b_{i}^{\dagger} = \sum_{p = 1}^{D} \frac{1}{2 \left( p - 1 \right) - E_{i}} c^{\dagger}_{p \uparrow} c^{\dagger}_{p \downarrow}
\end{equation}

\noindent and the parameters $E_{i}$ are the $D$ solutions of the equations

\begin{equation} \label{eqEi}
1 - \frac{g}{2} \sum_{p = 1}^{D} \frac{1}{2 \left( p - 1 \right) - E_i} - g \sum_{j \neq i }^{P} \frac{1}{E_i - E_j} = 0 \hspace{1cm} \forall i = 1, \dots, D
\end{equation}

The exact ground state energy is given by 

\begin{equation}
E_{GS} = \sum_{i = 1}^{P} E_i,
\end{equation}

\noindent where $E_i$ are the $P$ lowest energy solutions.

\subsection*{Basic DiagMC formalism}

Diagrammatic Monte Carlo (DiagMC) is based on the Dyson equation approach to the Green's function. The self-energy $\Sigma^{\star}$ is given by all one-particle irreducible Feynman diagrams\footnote{A diagram is said to be one-particle irreducible if it cannot be separated into two distinct diagrams by cutting one propagator line.}. In the self-consistent approach, the self-energy expansion is further reduced by including only skeleton diagrams, and hence DiagMC must sample one-particle irreducible skeleton diagrams. 
We build our DiagMC extension to the Richardson model building upon the formalism set in Ref. \cite{VanHoucke2019}.
To simplify the notation, we will assume spin-up unless otherwise specified, and define all one-body quantities as $O_{p} \defeq O_{p \uparrow, p \uparrow}$.

The self-energy is given by

\begin{equation} \label{eqDiagMC1}
\Sigma^{\star}_{p} \left( \omega \right) = \sum_{\mathcal{T}} \sum_{q_1, \dots, q_n} \int d \omega_1 \dots d \omega_m \mathcal{D}_p \left( \omega ; \mathcal{T}, q_1 \dots q_n, \omega_1 \dots \omega_m \right)   1_{\mathcal{T} \in S_{\Sigma^{\star}}},
\end{equation}

\noindent where $\mathcal{D}$ is a diagram with topology $\mathcal{T}$, internal single-particle quantum numbers $q_1\dots q_n$, and internal frequencies $\omega_1 \dots\omega_m$. The characteristic function $1_{\mathcal{T} \in S_{\Sigma^{\star}}}$ selects one-particle irreducible, skeleton topologies. Due to its analytic form, we expect the self-energy to be appreciably different from zero over a finite interval of energies $\omega$. We can divide this interval into bins and expand inside each bin $\left[ \omega_i, \omega_{i+1} \right]$ over a basis $\left\{ B_n \left( \omega \right) \right\}_{n=0}^{\infty}$.
We choose the complete orthonormal basis

\begin{equation} \label{eqBas}
B_n\left( \omega \right) = \sqrt{\frac{2n+1}{\omega_{i+1}-\omega_i}} L_n \left( \frac{2}{\omega_{i+1}-\omega_i}(\omega - \omega_i) - 1 \right) \; \; n = 0, 1, 2, \dots
\end{equation}

\noindent where $L_n \left( x \right)$ is the $n$-th Legendre polynomial. Since the order $n$ is also the number of nodes, a high-order expansion would introduce a strong source of sign problem. Accordingly, we truncate the expansion at $n_{max} = 2$ and adapt the size of for each bin to avoid significant oscillations inside the interval. From the exact analytic form of the self-energy we can estimate the width of the Lorentzians to be of the order of the regulator $\eta$. Hence, bins of dimension $\approx \eta$ ensure an accurate low-order truncation.
To avoid a large number of bins for small values of $\eta$, one can use an adaptive mesh in which the intervals near the Lorentzian peaks have size $\approx \eta$, while all other bins are much larger, even of the order of a few tens of energy units.
A preliminary computation with large intervals and few samples would suffice to estimate the pole position and the optimal mesh. For the present work, it wasn't necessary to adopt this refinement because the number of self-energy poles is limited\footnote{For $D=10$ there are five closeby poles.}. 

The self-energy expansion on the basis \eqref{eqBas} gives 

\begin{equation} \label{eqExpansion}
\Sigma_{p}^{\star} \left( \omega \right) = \sum_{n = 0}^{+ \infty} \Sigma_{p}^{(n)} B_n \left( \omega \right) \approx \sum_{n = 0}^{2} \Sigma_{p}^{(n)} B_n \left( \omega \right),
\end{equation}

\noindent with

\begin{equation} \label{eqSigmaCoeff}
\Sigma_{p}^{(n)} = \int_{\omega_i}^{\omega_{i+1}} d \omega \; B_n \left( \omega \right) \Sigma_{p}^{\star} \left( \omega \right).
\end{equation}

\noindent The DiagMC simulation calculates the coefficients \eqref{eqSigmaCoeff} inside each bin for every $p$ and allows reconstructing the self-energy through equation \eqref{eqExpansion}.

\noindent We can substitute \eqref{eqDiagMC1} into \eqref{eqSigmaCoeff}, and using the shorthand notation $\mathcal{C} \defeq \left( \mathcal{T}, q_1, \dots, q_n, \omega_1, \dots, \omega_m \right) $ we get

\begin{equation} \label{eqDiagMC2}
\Sigma_{p}^{(n)} =  \int_{\omega_i}^{\omega_{i+1}} d \omega \int d \mathcal{C} \; B_n \left( \omega \right) \mathcal{D}_p \left( \omega; \mathcal{C} \right) 1_{\mathcal{T} \in S_{\Sigma^{\star}}}.
\end{equation}

The main idea of DiagMC is to rewrite this expansion as a weighted average over a probability distribution function $w_{p}$. Choosing $w_{p}$ to be independent of $n$ allows us to evaluate the coefficients $\Sigma_p^{(n)}$ for every $n$ with a single simulation instead of performing separate samplings for each $n$ up to $n_{max}$. We still require a different simulation for every basis-state level $p = 1, \dots, D$. To this end, we rewrite \eqref{eqDiagMC2} as

\begin{equation} \label{eqDiagMC3}
\Sigma_{p}^{(n)} = Z_p \int_{\omega_i}^{\omega_{i+1}} d \omega \int d \mathcal{C} \; \frac{\lvert \mathcal{D}_p \left( \omega; \mathcal{C} \right) \rvert}{Z_p} \exp \left[ i \arg\mathcal{D}_p \left( \omega; \mathcal{C} \right) \right] B_n \left( \omega \right) 1_{\mathcal{T} \in S_{\Sigma^{\star}}},
\end{equation}

\noindent where $Z_p$ is a normalization factor given by

\begin{equation} \label{eqNormFact}
Z_p \defeq \int_{\omega_i}^{\omega_{i+1}} d \omega \int d \mathcal{C} \; \lvert \mathcal{D}_p \left( \omega; \mathcal{C} \right) \rvert.
\end{equation}

\noindent We can define the probability distribution function as

\begin{equation} \label{eqProbDist}
w_p \left(\omega; \mathcal{C} \right) \defeq \frac{ \lvert \mathcal{D}_p \left( \omega; \mathcal{C} \right) \rvert }{Z_p}
\end{equation}

\noindent and rewrite Eq.~\eqref{eqDiagMC3} as 

\begin{equation} \label{eqDiagMC4}
\Sigma_{p}^{(n)} = Z_p \int_{\omega_i}^{\omega_{i+1}} d \omega \int d \mathcal{C} \; w_p \left(\omega; \mathcal{C} \right)  \exp \left[ i \arg\mathcal{D}_p \left( \omega; \mathcal{C} \right) \right] B_n \left( \omega \right) 1_{\mathcal{T} \in S_{\Sigma^{\star}}}.
\end{equation}

\noindent If we can sample diagrams $\left\{ \mathcal{D}_p \left( \omega_j; \mathcal{C}_j \right)\right\}_{j = 1}^N$ according to the probability distribution function \eqref{eqProbDist}, we can calculate \eqref{eqDiagMC4} with

\begin{equation} \label{eqDiagMCFond1} 
\Sigma_{p}^{(n)} = Z_p \lim_{N \to \infty} \frac{1}{N} \sum_{j=1}^{N} \exp \left[ i \arg\mathcal{D}_p \left( \omega_j; \mathcal{C}_j \right) \right] B_n \left( \omega_j \right) 1_{\mathcal{T}_j \in S_{\Sigma^{\star}}}.
\end{equation}

The normalization factor $Z_p$ is estimated during the DiagMC simulation by selecting a subset of diagrams $S_{\mathcal{N}}$, called normalization sector, with a known weight $Z^{\mathcal{N}}_{p}$. Since

\begin{equation} \label{eqZN} 
Z^{\mathcal{N}}_{p} \defeq \int_{\omega_i}^{\omega_{i+1}} d \omega \int_{\mathcal{T} \in S_{\mathcal{N}} } d \mathcal{C} \; \lvert \mathcal{D}_p \left( \omega; \mathcal{C} \right) \rvert,
\end{equation}

\noindent we can estimate the normalization factor by counting the number of times these diagrams are sampled. If the number of visits to the normalization sector is $\mathcal{N}$, the probability distribution function \eqref{eqProbDist} implies

\begin{equation} \label{eqDiagFond2}
\lim_{N \to \infty} \frac{\mathcal{N}}{N} = \frac{Z^{\mathcal{N}}_{p}}{Z_p}
\end{equation}

\noindent and Eq.~\eqref{eqDiagMCFond1} becomes

\begin{equation} \label{eqDiagMCFond3} 
\Sigma_{p}^{(n)}= Z^{\mathcal{N}}_{p}\lim_{N \to \infty} \frac{1}{\mathcal{N}} \sum_{j=1}^{N} \exp \left[ i \arg\mathcal{D}_p \left( \omega_j; \mathcal{C}_j \right) \right] B_n \left( \omega_j \right) 1_{\mathcal{T}_j \in S_{\Sigma^{\star}}}.
\end{equation}

We discuss the normalization sector in depth in the next subsection.

\subsection*{Hartree-Fock and normalization sector}

\begin{figure}[ht]
\centering
\includegraphics[scale=1]{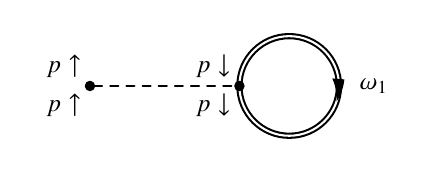}
\caption{First-order (tadpole) diagram. This diagram accounts for the static part of the self-energy~\cite{Dickhoff:manybody}.}
\label{figHF}
\end{figure}

The Hartree-Fock diagram in Fig. \ref{figHF} has the analytic expression

\begin{equation} \label{eqInfProb}
\mathcal{D}^{\text{HF}}_p = i \frac{g}{2}  \int_{-\infty}^{\infty} \frac{d\omega_1}{2 \pi} G_{p \downarrow p \downarrow} \left( \omega_1 \right) e^{i \omega_1 \epsilon} = i \frac{g}{2}  \int_{-\infty}^{\infty} \frac{d \omega_1}{2 \pi} G_{p} \left( \omega_1 \right) e^{i \omega_1 \epsilon},
\end{equation}

\noindent where the last equality is due to the spin-up - spin-down symmetry.

Eq.~\eqref{eqInfProb} is difficult to calculate with DiagMC due to the infinitesimal $\epsilon$ that needs to be taken to $0^+$ after the integral. Thus, we compute the result analytically from

\begin{equation} \label{eqHF}
\mathcal{D}^{\text{HF}}_p = -\frac{g}{2} \sum_k \lvert
\langle \Psi_k^{2P-1} | c_p| \Psi_{g.s.}^{2P} \rangle
\rvert^2.
\end{equation}

Eq.~\eqref{eqHF} is the total static self-energy. To avoid the problems of Eq.~\eqref{eqInfProb}, we calculate the static self-energy exactly through Eq.~\eqref{eqHF} and simulate the dynamic self-energy with DiagMC. This is equivalent to considering $S_{\Sigma^{\star}}$ in \eqref{eqDiagMCFond3} as the subset of skeleton diagrams of order greater than $1$.

\begin{figure}[ht]
\centering
\includegraphics[scale=1]{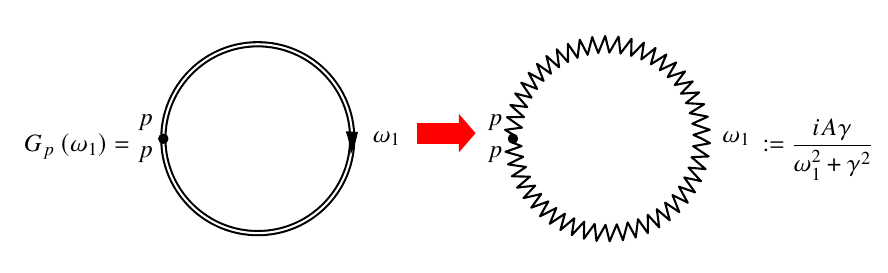}
\caption{Self-closing physical and unphysical propagators. The self-closing physical propagator (represented by a double solid line) is replaced with an unphysical propagator (represented by a zigzag line) with an arbitrarily chosen value.}
\label{figSelfClosing}
\end{figure}

Since the Hartree-Fock diagram has been removed from the set of sampled diagrams, and we include only skeleton diagrams in the self-energy simulation, we can redefine the self-closing propagator to

\begin{equation} \label{eqClosedProp}
\tilde{G}_p \left( \omega_1 \right) = \frac{i A \gamma}{\omega_1^2 + \gamma^2},
\end{equation}

\noindent with $A, \gamma > 0$. The redefined self-closing propagator will be drawn as a zigzag line, as shown in Fig. \ref{figSelfClosing}.
This substitution allows us to take the limit $\epsilon \to 0^+$ before performing the integral.

The redefinition of the HF diagram allows the manipulation of the weight (and hence the amount of sampling) of the diagram. This is a useful feature to keep under control the normalization sector. Hence, we take as normalization sector $S_{\mathcal{N}}$ the HF diagram in Fig. \ref{figHF}, with the closed propagator replaced by the unphysical zigzag propagator \eqref{eqClosedProp}. The weight of the normalization sector is found analytically from Eq.~\eqref{eqZN} 

\begin{equation}
Z^{\mathcal{N}}_p = \frac{\lvert g \rvert}{2} \int_{\omega_i}^{\omega_{i+1}} d \omega \int_{- \infty}^{+ \infty} \frac{d \omega_1}{2 \pi} \frac{A \gamma}{\omega_1^2 + \gamma^2} = \frac{\lvert g \rvert}{4} A \Delta \omega \defeq Z^{\mathcal{N}},
\end{equation}

\noindent which is independent of $p$ and where $\Delta \omega \defeq \omega_{i + 1} - \omega_{i}$. The parameters $A$ and $\gamma$ can be adjusted to optimize the sampling of normalization diagrams. It was sufficient to choose $\gamma \approx 15$ and $A \approx 3$ throughout this work, although they could be fine-tuned according to the coupling constant $g$, the model space dimension, and the order of the perturbative expansion.

\subsection*{Updates}

Eq.~\eqref{eqDiagMCFond3} implies sampling the space of Feynman diagram topologies. While this includes all skeleton diagrams of order greater than one, the diagrammatic expansion of the Richardson model is dominated by ladder diagrams. Here, we limit ourselves to this subset and sample diagrams up to order eight. The same algorithm can be pushed to higher-order contributions without modifications. We construct a Markov chain that explores the space of topologies, quantum numbers, and frequencies, using $10^3$ thermalization steps to randomize the initial state.

The following diagrams are proposed by randomly choosing one of five possible updates that modify the diagram at the previous step. 
To reproduce the correct probability distribution function \eqref{eqProbDist}, we accept or reject the update according to Metropolis-Hastings algorithm~\cite{Metropolis1953, Hasting1970}. 
The updates we used to sample ladder contributions are as follows:

\begin{itemize}
    \item \emph{Change of} $\omega$

    This update changes the external frequency $\omega$ of the diagram. To preserve frequency conservation, we change the frequency of the backward-going propagator as in Fig. \ref{figChangeOmega}.

    \begin{figure}[ht]
    \centering
    \includegraphics[scale=0.95]{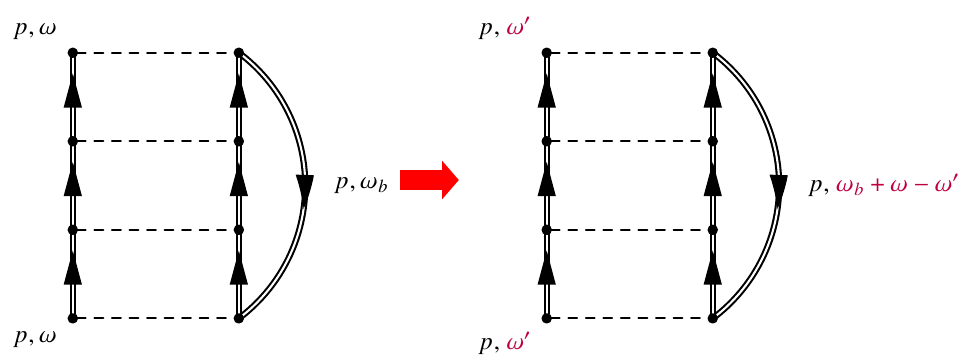}
    \caption{Update \emph{Change} $\omega$.}
    \label{figChangeOmega}
    \end{figure}
    
    The new frequency $\omega'$ is chosen according to a uniform distribution in the bin $\left[ \omega_i, \omega_{i+1} \right]$ and the Metropolis-Hastings acceptance ratio is

    \begin{equation}
    q_{\omega} = \frac{\lvert G_p \left( \omega_{b} + \omega - \omega' \right) \rvert}{\lvert G_p \left( \omega_{b}  \right) \rvert}.
    \end{equation}

    \item \emph{Change of the internal frequencies}

    This update changes the frequency $\omega_b$ of the backward-going propagator that connects the two ends of the ladder. We sample new frequencies, both in this update and in \emph{Change of single-particle quantum numbers and frequencies}, following a Lorentzian distribution $L(\omega)$. This choice is motivated by general considerations on importance sampling.  
    Focusing on any particular internal frequency $\omega_i$, the number of propagators carrying $\omega_i$ and the K\"{a}ll\'{e}n-Lehmann representation ensure that the diagram decays as $1/\omega_i^l$, with $l \geq 2$. To minimize the variance of the integral over $\omega_i$, we choose a Lorentzian distribution due to its inverse polynomial decay $1/\omega_i^2$, which closely reproduces the asymptotic behavior of internal frequencies in the diagram. The sampling of bulk frequencies, where propagators assume large values, can be controlled by adjusting the Lorentzian width. We found that any width of $\mathcal{O}(10)$ was sufficient for fast convergence.
    The propagators on the left-hand side of the ladder diagram have their frequency $\omega_i$ changed to $ \omega_i - \omega_b + \omega_b'$. The update is shown in Fig. \ref{figChangeOmegaInt}.

    \begin{figure}[ht]
    \centering
    \includegraphics[scale=.94]{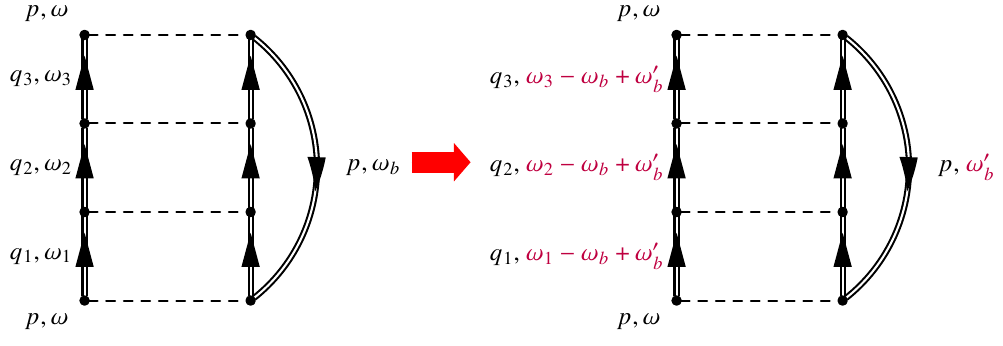}
    \caption{Update \emph{Change of the internal frequencies}.}
    \label{figChangeOmegaInt}
    \end{figure}

    The acceptance ratio of the update is 

    \begin{equation}
    q_{\omega \; int} = \frac{L ( \omega_{b} )}{L ( \omega_{b}')} \frac{\lvert G_p \left( \omega_{b}' \right) \rvert}{\lvert G_p \left( \omega_{b} \right) \rvert} \prod_{j = 1}^{\text{order}-1} \frac{\lvert G_{q_j} \left( \omega_j - \omega_{b} + \omega_{b}' \right) \rvert}{\lvert G_{q_j} \left( \omega_j \right) \rvert}.
    \end{equation}

    Notice that this is the only update that can move over the entire real axis the frequency of the backward-going propagator connecting the two ends of the ladder. This is in contrast to the update \emph{Change $\omega$}, which can only change $\omega_b$ by an amount within the interval $\left[\omega_i, \omega_{i+1} \right]$. This update is needed to perform the integration over $\omega_b$.

    \item \emph{Change of single particle quantum numbers and frequencies}

    This update chooses a random propagator on the left-hand side of the ladder and proposes to change its frequency and single-particle quantum number. The new frequency is sampled from the same Lorentzian distribution of the update \emph{Change of the internal frequencies}. The new single-particle quantum number is drawn from a uniform probability distribution on the integers $\left[ 1, D \right]$. The propagator on the right-hand side has its frequency changed to conserve the total frequency. The single-particle quantum number on the right-hand side is also changed to match the one on the left, as required by the particular form of the interaction.

    \begin{figure}[ht]
    \centering
    \includegraphics[scale=.95]{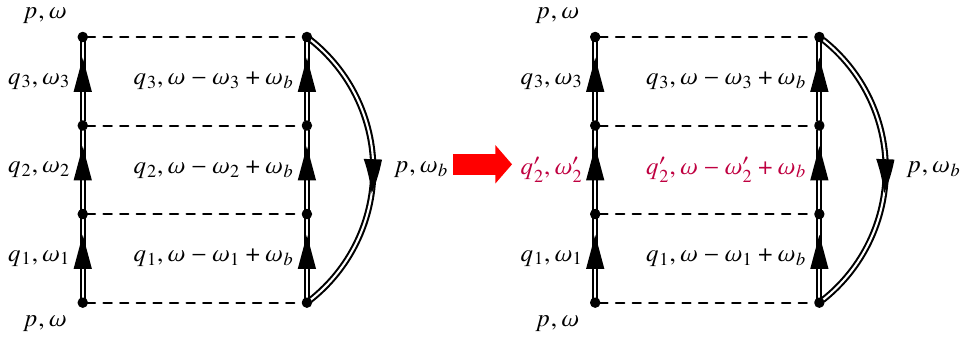}
    \caption{Update \emph{Change single particle quantum numbers and frequencies}. We assumed that the random propagator chosen was the second from the bottom appearing on the left-hand side of the ladder.}
    \label{figdiagram_ChangeOmegaSpQN}
    \end{figure}

    This update is depicted in Fig. \ref{figdiagram_ChangeOmegaSpQN} and its ratio is

    \begin{equation}
    q_{q, \omega} = \frac{L \left( \omega_{2} \right)}{L ( \omega_{2}' )} \frac{\lvert G_{q_2'} \left( \omega_2' \right) G_{q_2'} \left(\omega - \omega_2' + \omega_b \right) \rvert}{\lvert G_{q_2} \left( \omega_2 \right) G_{q_2} \left(\omega - \omega_2 + \omega_b \right) \rvert}.
    \end{equation}

    \item \emph{Add/Remove Rung}

    The \emph{Add/Remove Rung} updates are a pair of complementary updates that propose to add (remove) a pair of propagators to (from) the top of the diagram. When called on the normalization diagram, the \emph{Add Rung} update proposes to turn it into the second-order ladder, while \emph{Remove Rung} is always rejected. When \emph{Remove Rung} is called upon the second-order diagram, it proposes returning to the normalization sector. Accordingly, the backward-going propagator that connects the ends of the ladder is changed into the unphysical propagator of Fig. \ref{figSelfClosing}, and vice-versa when moving to the second-order ladder diagram.
    The simulations are performed with an order cutoff implemented by rejecting the \emph{Add Rung} update when called on diagrams of the maximum order. We show the update in Fig. \ref{figAddRemLadderAdv}. The frequency on the left-hand side of the top pair is drawn from the same Lorentzian function $L \left( \omega \right)$ of the other updates, and the frequency of the right propagator is adjusted to preserve frequency conservation. The single-particle quantum number is chosen randomly among the integers $\left[1, D \right]$.

    \begin{figure}
    \centering
    \includegraphics[scale=1]{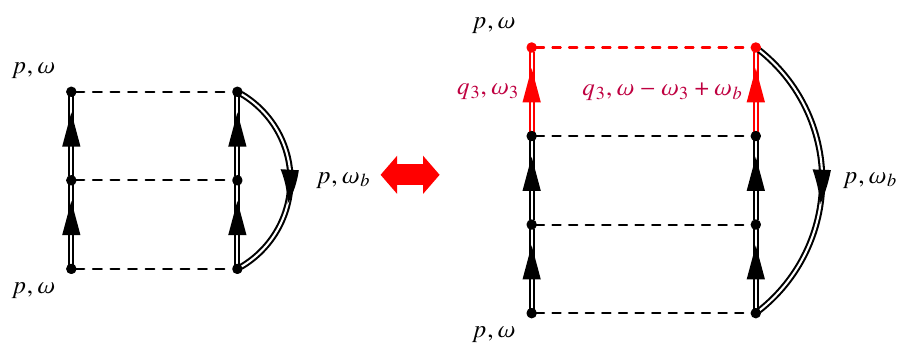}
    \caption{Update \emph{Add/Remove Rung}. The example in the picture shows the second and third-order ladder diagrams.}
    \label{figAddRemLadderAdv}
    \end{figure}

\noindent The update ratio of the \emph{Add Rung} depicted in Fig. \ref{figAddRemLadderAdv} is 

\begin{equation}
q_{AR} = \frac{\lvert g \rvert}{4 \pi} \frac{D}{L \left( \omega_3 \right)} \lvert G_{q_3} \left( \omega_3 \right) G_{q_3} \left( \omega - \omega_3 + \omega_b \right) \rvert,
\end{equation}

\noindent while for the \emph{Remove Rung} it is

\begin{equation}
q_{RR} = \frac{1}{q_{AR}}.
\end{equation}

\end{itemize}

The updates presented can sample all ladder diagrams ergodically up to arbitrarily high orders. 

\subsection*{Dyson matrix diagonalization}

DiagMC calculates the coefficients \eqref{eqSigmaCoeff} of the self-energy expansion, Eq.~\eqref{eqExpansion}. The causality principle constrains the full form of the dynamic self-energy by its imaginary part. Consequently, we can focus on the imaginary part of the self-energy and perform a non-linear least-square minimization to fit the imaginary part of the self-energy as a sum of Lorentzian functions. 
The functional form of the fit is motivated by the exact analytic form of the self-energy given by the K\"{a}llén-Lehmann representation. If the coupling constant lies within the convergence radius of the series expansion, the perturbative expansion must approach the analytic form of the exact self-energy. We focus on this small coupling regime. 
If the regulator $\eta$ of the propagator is finite, the imaginary part of the self-energy goes from being a sum of Dirac $\delta$ functions~\cite{Dickhoff:manybody} to a sum of Lorentzian functions with finite width,

\begin{equation}
\text{Im} \left[ \Sigma_p^{\star} \left( \omega \right) \right] = \sum_{j=1}^N \frac{\pm C_p^jA_p^j}{\left( \omega - B_p^j \right)^2 + {C_p^j}^2},
\end{equation}

\noindent where $A_p^j, C_p^j > 0$, the overall sign is $+$ ($-$) for hole (particle) poles and the number  $N$ is clearly determined by the shape of the self-energy (see Fig. \ref{figIncrOrd}). Each Lorentzian has three free parameters: the strength $A_p^j$, the centroid $B_p^j$, and the width $C_p^j$. This set of parameters, combined with the static self-energy given by the HF diagram, gives the full self-energy

\begin{equation}
\Sigma_{p}^{\star} \left( \omega \right) = \Sigma_{p}^{\infty} + \sum_{j=1}^N \frac{A_p^j}{\omega - B_p^j \mp i C_p^j}.
\end{equation}

This form lets us take the limit $C_p^j \to 0^+$. The dependence on the regulator is suppressed in the explicit expression of the self-energy, however, it is still hidden in the parameters $A_p^j$ and $B_p^j$.  

The exact analytic form of the propagator is 

\begin{equation}
G_{p} \left( \omega \right) = \sum_j \frac{\lvert \mathcal{Z}_p^j \rvert^2}{\omega - \varepsilon_j^{\pm} \pm i \eta},
\end{equation}

\noindent where $\mathcal{Z}_p^j$ are the overlap amplitudes for quasiparticle ($\varepsilon_j^{+}$) and quasihole ($\varepsilon_j^{-}$) solutions.

We can find the poles and residues of the propagator by solving the Dyson equation. This equation can be recast into a standard eigenvalue problem \cite{Barbieri::LectNotesPhys936}. We can further exploit the diagonal nature of the propagator to divide the eigenvalue problem into $D$ smaller eigenvalue problems of the form

\begin{equation} \label{eqDysMat}
    \begin{pmatrix}
        p - 1 + \Sigma_{p}^{\infty} & \sqrt{A_p^1} & \sqrt{A_p^2} & \dots & \sqrt{A_p^N} \\
        \sqrt{A_p^1} & B_p^1 & 0 & \dots & 0 \\
        \sqrt{A_p^2} & 0 & B_p^2 & \dots & 0 \\
        \vdots & \vdots & \vdots & \ddots & \vdots \\
        \sqrt{A_p^N} & 0 & 0 & \dots & B_p^N
    \end{pmatrix} 
    \begin{pmatrix}
    \mathcal{Z}_p^j \\
    W^1 \\
    W^2 \\
    \vdots \\
    W^N
    \end{pmatrix} =
    \varepsilon_j^{\pm} 
    \begin{pmatrix}
    \mathcal{Z}_p^j \\
    W^1 \\
    W^2 \\
    \vdots \\
    W^N
    \end{pmatrix}
\end{equation}

\noindent with $p=1,\dots, D$. To determine whether a pole is particle-like or hole-like we look at the position of its energy $\varepsilon_j^{\pm}$ with respect to the Fermi energy. 
We compute the Fermi energy by requiring that the total sum of spectroscopic factors over the quasihole states is equal to the number of particles. 

\noindent Eq.~\eqref{eqDysMat}, complemented with the normalization condition 

\begin{equation} \label{eqNormCond}
   \lvert \mathcal{Z}_p^j \rvert^2 + \sum_{k = 1}^N \lvert W^k \rvert^2 = 1,
\end{equation}

\noindent fully determines the propagator.

\subsection*{Convergence of the perturbation series}

The perturbation series diverges outside of the convergence radius of the expansion, $g_c$. This radius increases when employing larger values of the regulator $\eta$, which implies lowering the energy resolution. For the adopted resolution ($\eta = 0.1$), the radius of convergence of the perturbation theory expansion is $\gtrsim 0.3$. Outside of this range, the series diverges, with the consequence of breaking the causality structure.

\begin{figure*}[ht]
\centering
\includegraphics[scale = 0.47]{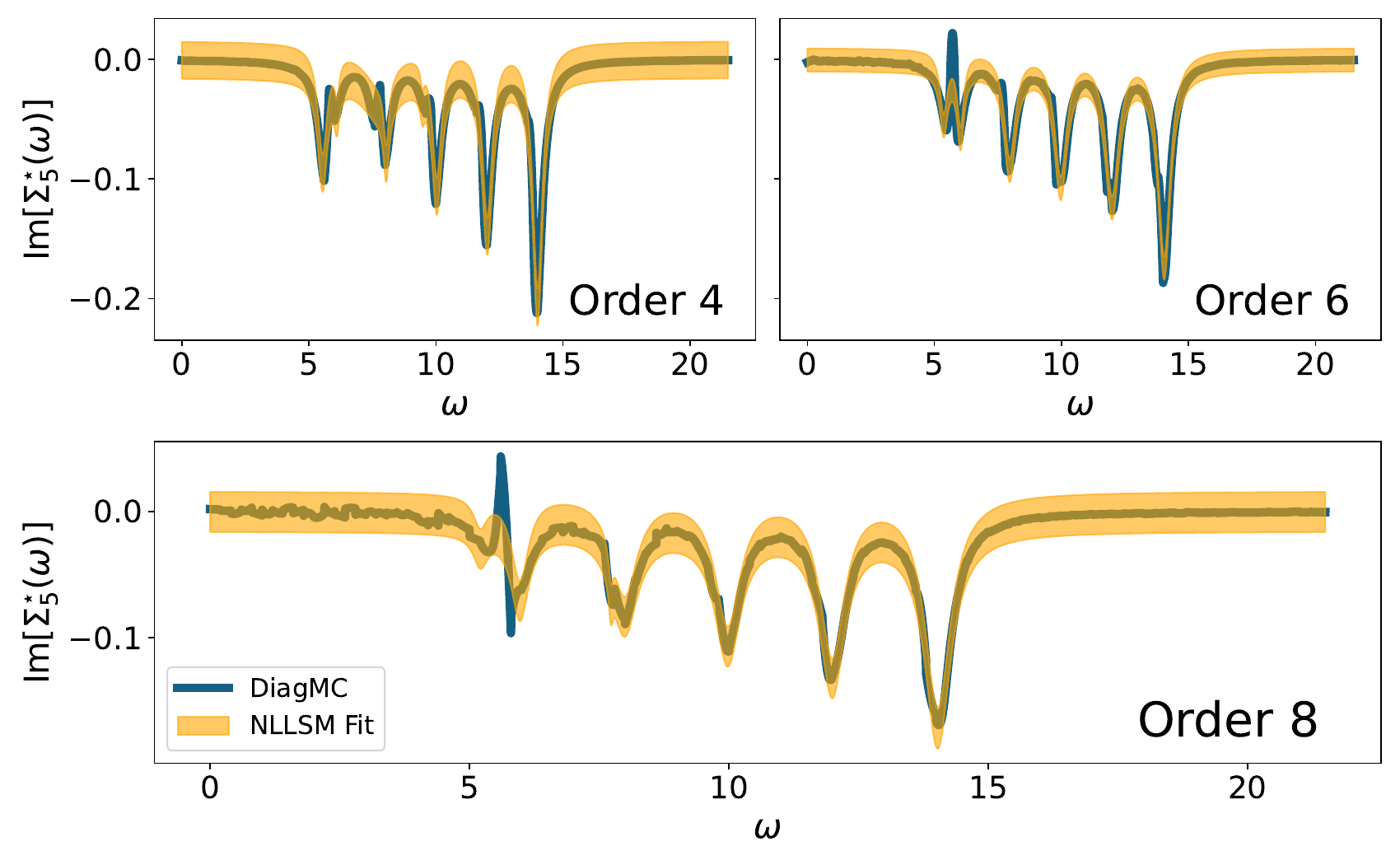}
\caption{Imaginary part of the component $p = 5$ of the self-energy calculated at order $4$, $6$, and $8$ in the ladder expansion and $g = -0.4$, $\eta = 0.1$.}
\label{figRadConv}
\end{figure*}

Fig.~\ref{figRadConv} demonstrates the  $p=5$ component of the imaginary part of the self-energy for $g=-0.4$ and $\eta = 0.1$ and it exhibits an anomalous spike at $\omega \sim 5.5$, in the proximity of one of the poles. This spike persists (and becomes more pronounced) when increasing the order of the ladder simulation. This pole and spike structure is characteristic of this type of divergence and it is common to methods based on perturbation theory expensions. It presents itself also for larger values of $\lvert g \rvert$.

\begin{figure}[ht]
\centering
\includegraphics[scale = .7]{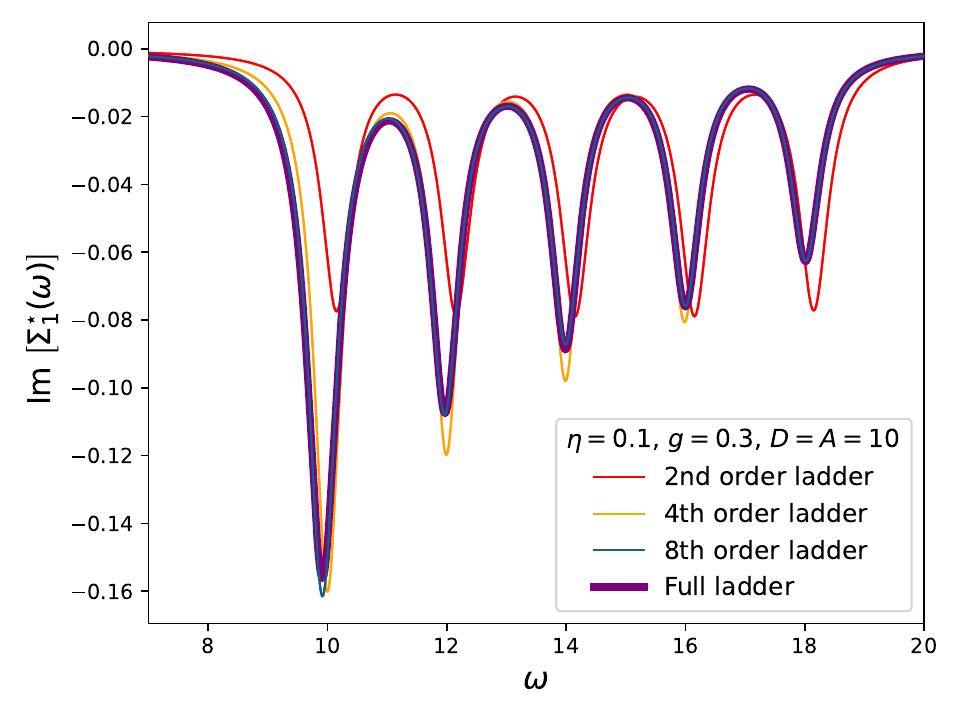}
\caption{Imaginary part of the component $p=1$ of the self-energy. The values of the parameters of the calculation are  $g=0.3$, $\eta=0.1$ and $D=10$.}
\label{figIncrOrd}
\end{figure}

To determine the convergence of the ladder expansion with respect to the truncation order, we computed the exact ladder resummation, and then we performed DiagMC calculations at increasing order. Fig. \ref{figIncrOrd} shows the imaginary part of the component $p=1$ of the self-energy. The self-energy was calculated in the model space of size $D = 10$ for the coupling constant $g=0.3$ and regulator $\eta = 0.1$. The perturbation expansion is approaching the exact resummation as the order increases, and it reaches convergence at the 8${}^{\rm{th}}$ order.

\subsection*{Convergence of MCMC and error determination}

We divide the entire energy interval into bins $[\omega_i, \omega_{i+1}]$ of size $\sim \eta$ and perform Monte Carlo Markov chain simulations independently for each bin.
For the present purposes, it is sufficient to sample with a single walker per bin. Each Markov chain starts with $10^3$ thermalization steps, after which up to $5\times10^{10}$ steps are collected for statistics.

\begin{figure*}[ht]
\centering
\includegraphics[scale = 0.545]{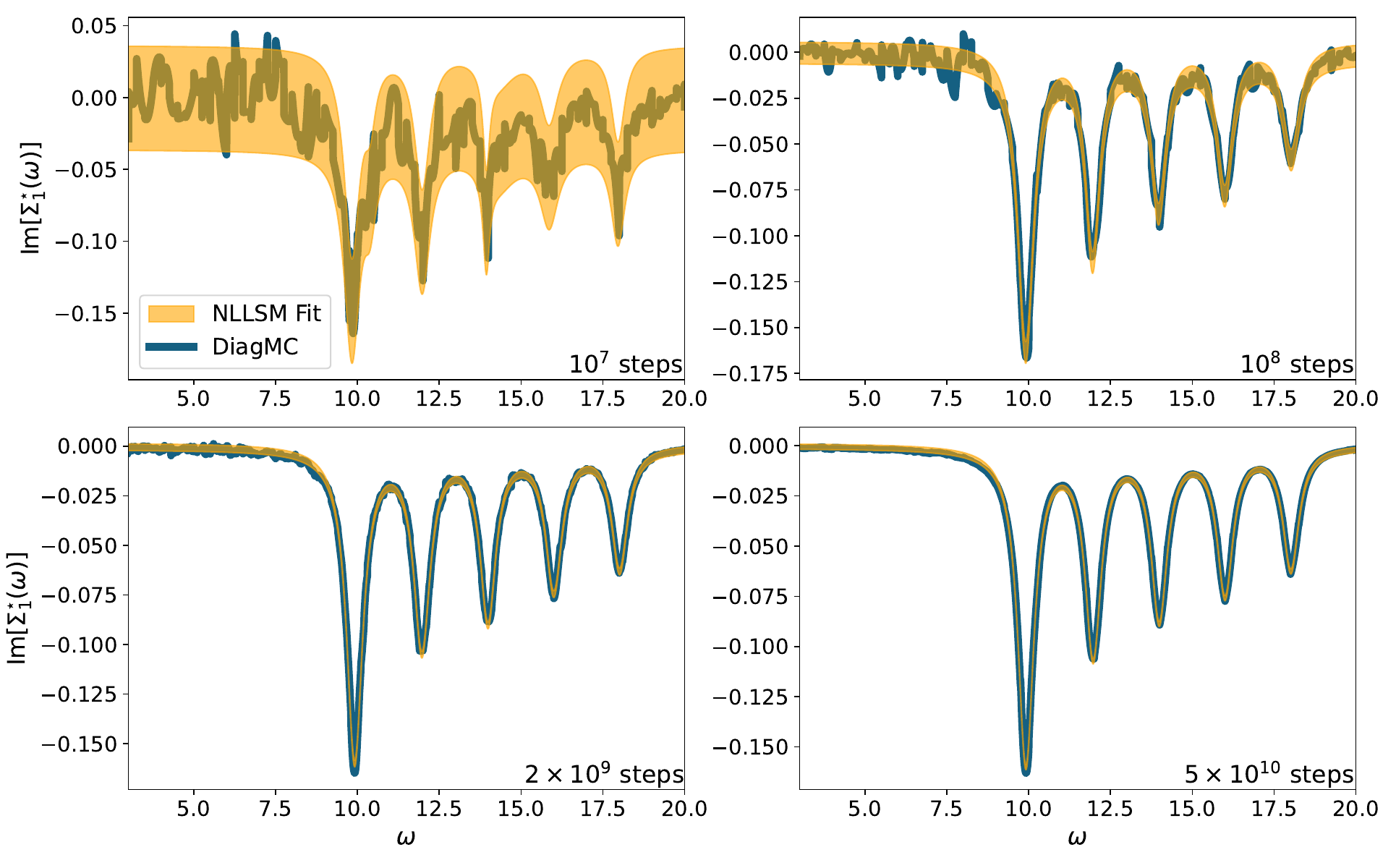}
\caption{Imaginary part of the component $p=1$ of the self-energy for $g = 0.3$ and $\eta = 0.1$. Calculations with increasing numbers of steps per walker are shown. The figures show both the row data of DiagMC simulations and the NNLSM fits as sums of Lorentzians.}
\label{figConvMCMC}
\end{figure*}

\begin{figure*}[ht]
\centering
\includegraphics[scale = 0.545]{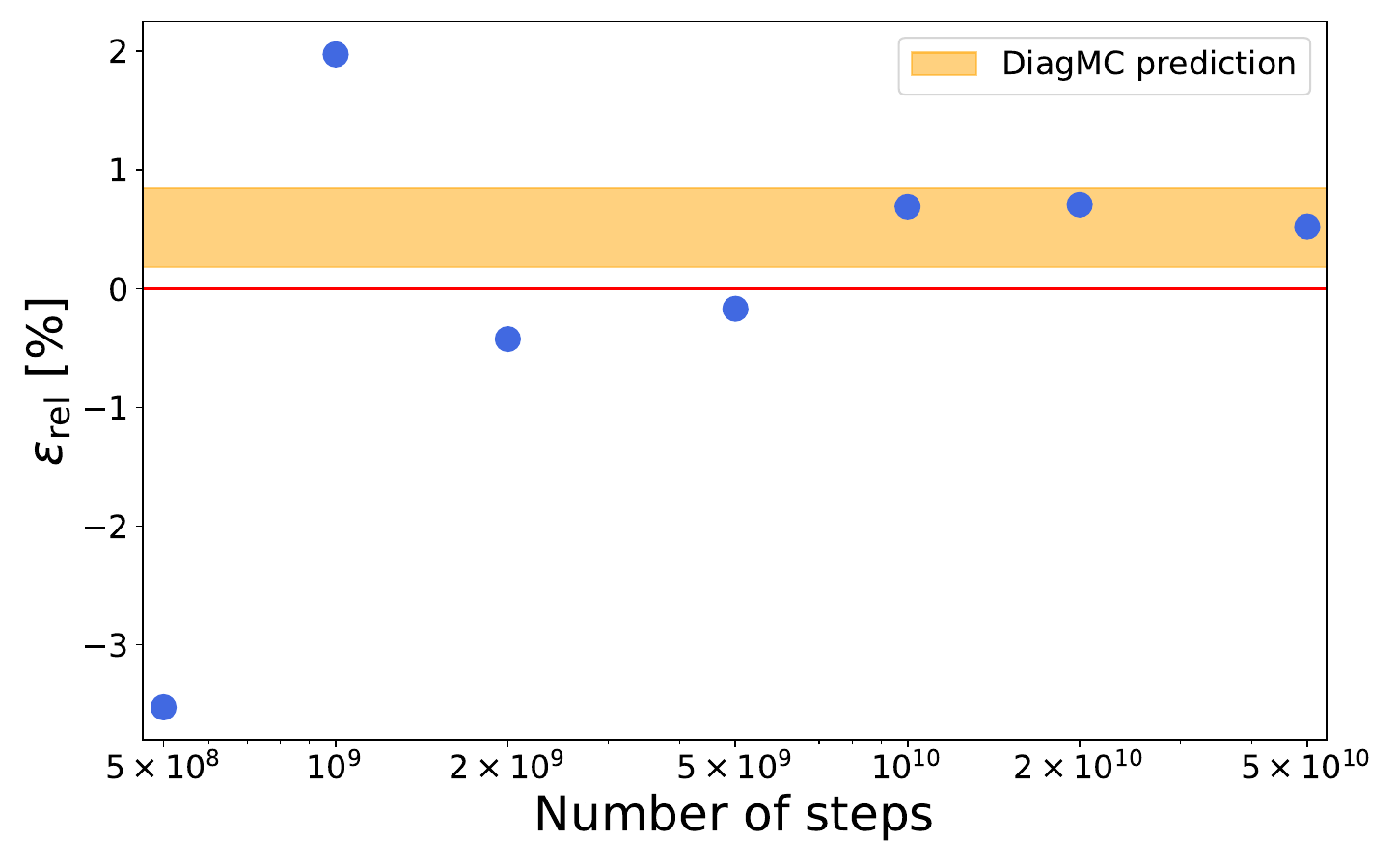}
\caption{Percentage deviation of the correlation energy as a functions of the number of steps per walker. For each point, a new simulation was performed from scratch independently (without reusing data from larger samples).
The orange band is the weighted standard error around the weighted average of the simulations shown in the plot.}
\label{figEcorr_steps}
\end{figure*}

Fig.~\ref{figConvMCMC} illustrates the convergence with respect to the total number of MCMC steps for the first component of the imaginary part of the self-energy. The blue lines are the raw MC data, while the yellow bands are the corresponding NLLSM fits. The widths of each band are the uncertainties propagated to the fitted parameters $A_p^{n/k}$ and $B_p^{n/k}$. Convergence is largely achieved at approximately $\sim 10^9$ steps, but we were able to sample an order of magnitude more to completely eliminate this source of uncertainty.
A full simulation entails sampling all components of the self-energy along the range of frequencies \hbox{$\omega\in[-15,21]$.} Following this, the total energy can be computed easily by solving the Dyson equation and applying the Migdal-Galitskii-Koltun sum rule. Fig.~\ref{figEcorr_steps} shows the percentage deviation of the final correlation energy with respect to the exact result $\varepsilon_{\rm{rel}} = (E_{\rm{corr}}^{\rm{DiagMC}} -  E_{\rm{corr}}^{\rm{exact}})/\lvert E_{\rm{corr}}^{\rm{exact}} \rvert$. This includes statistical MCMC uncertainty, the truncation of the diagrammatic series to ladders up to the eighth order, and possible uncertainties arising from the choice of the finite regulator, $\eta=0.1$. Note that all simulations with more than $2 \times 10^9$ steps agree within $1\%$ of each other and the exact result. The orange band represents the weighted standard error around the weighted average of the simulations displayed in the plot.

\clearpage
\bibliographystyle{apsrev4-2}
\bibliography{bibliography}